\documentclass[twocolumn,letterpaper,aps,prc,superscriptaddress,showpacs,floatfix]{revtex4}

\usepackage{graphicx}	
\usepackage{xspace}	

\newcommand{\pt}{\mbox{$p_T$}\xspace}

\newcommand{\Npart}{\mbox{$N_{\rm part}$}\xspace}
\newcommand{\Ncoll}{\mbox{$N_{\rm coll}$}\xspace}

\newcommand{\mNcoll}{\mbox{$\langle N_{\rm coll} \rangle$}\xspace}
\newcommand{\sqs}{\mbox{$\sqrt{s}$}\xspace}
\newcommand{\sqsn}{\mbox{$\sqrt{s_{_{NN}}}$}\xspace}
\newcommand{\dau}{\mbox{$d$$+$Au}\xspace}
\newcommand{\pp}{\mbox{$p$$+$$p$}\xspace}
\newcommand{\ppb}{\mbox{$p$$+$Pb}\xspace}

\newcommand{\rdau}{\mbox{$R_{dAu}$}\xspace}
\newcommand{\pda}{\mbox{$p(d)$+A}\xspace}
\newcommand{\rpda}{\mbox{$R_{p(d)+A}$}\xspace}
\newcommand{\bbceta}{\mbox{$3.0<|\eta|<3.9$}\xspace}

\begin{document}


\title{Centrality categorization for $R_{p(d)+A}$ in high-energy collisions}

\date{\today}

\
\begin{abstract}

High energy proton- and deuteron-nucleus collisions provide an excellent 
tool for studying a wide array of physics effects, including modifications 
of parton distribution functions in nuclei, gluon saturation, and color 
neutralization and hadronization in a nuclear environment, among others.  
All of these effects are expected to have a significant dependence on the 
size of the nuclear target and the impact parameter of the collision, also 
known as the collision centrality.  In this article, we detail a method for 
determining centrality classes in $p(d)+A$ collisions via cuts on the 
multiplicity at backward rapidity (i.e., the nucleus-going direction) and 
for determining systematic uncertainties in this procedure.  For $d$$+$Au 
collisions at $\sqrt{s_{_{NN}}}=200$~GeV we find that the connection to 
geometry is confirmed by measuring the fraction of events in which a 
neutron from the deuteron does not interact with the nucleus.  As an 
application, we consider the nuclear modification factors $R_{p(d)+A}$,
for which there is a bias in the measured centrality-dependent 
yields due to auto correlations between the process of interest and the 
backward-rapidity multiplicity.  We determine the bias-correction factors 
within this framework.  This method is further tested using the {\sc hijing}
MC generator.  We find that for $d$$+$Au collisions at 
$\sqrt{s_{_{NN}}}=200$~GeV, these bias corrections are small and vary by 
less than 5\% (10\%) up to $p_{T}$=10~(20)~GeV/$c$.  In contrast, for 
$p$$+$Pb collisions at $\sqrt{s_{_{NN}}}=5.02$~TeV we find these bias 
factors are an order of magnitude larger and strongly $p_T$ dependent, 
likely due to the larger effect of multiparton interactions.

\end{abstract}

\newcommand{\abilene}{Abilene Christian University, Abilene, Texas 79699, USA}
\newcommand{\augie}{Department of Physics, Augustana College, Sioux Falls, South Dakota 57197, USA}
\newcommand{\banaras}{Department of Physics, Banaras Hindu University, Varanasi 221005, India}
\newcommand{\barc}{Bhabha Atomic Research Centre, Bombay 400 085, India}
\newcommand{\baruch}{Baruch College, City University of New York, New York, New York, 10010 USA}
\newcommand{\bnlcoll}{Collider-Accelerator Department, Brookhaven National Laboratory, Upton, New York 11973-5000, USA}
\newcommand{\bnlphys}{Physics Department, Brookhaven National Laboratory, Upton, New York 11973-5000, USA}
\newcommand{\caucr}{University of California - Riverside, Riverside, California 92521, USA}
\newcommand{\charlesczech}{Charles University, Ovocn\'{y} trh 5, Praha 1, 116 36, Prague, Czech Republic}
\newcommand{\chonbuk}{Chonbuk National University, Jeonju, 561-756, Korea}
\newcommand{\ciae}{Science and Technology on Nuclear Data Laboratory, China Institute of Atomic Energy, Beijing 102413, P.~R.~China}
\newcommand{\cns}{Center for Nuclear Study, Graduate School of Science, University of Tokyo, 7-3-1 Hongo, Bunkyo, Tokyo 113-0033, Japan}
\newcommand{\colorado}{University of Colorado, Boulder, Colorado 80309, USA}
\newcommand{\columbia}{Columbia University, New York, New York 10027 and Nevis Laboratories, Irvington, New York 10533, USA}
\newcommand{\czechtech}{Czech Technical University, Zikova 4, 166 36 Prague 6, Czech Republic}
\newcommand{\dapnia}{Dapnia, CEA Saclay, F-91191, Gif-sur-Yvette, France}
\newcommand{\elte}{ELTE, E{\"o}tv{\"o}s Lor{\'a}nd University, H - 1117 Budapest, P{\'a}zm{\'a}ny P. s. 1/A, Hungary}
\newcommand{\ewha}{Ewha Womans University, Seoul 120-750, Korea}
\newcommand{\fit}{Florida Institute of Technology, Melbourne, Florida 32901, USA}
\newcommand{\fsu}{Florida State University, Tallahassee, Florida 32306, USA}
\newcommand{\gsu}{Georgia State University, Atlanta, Georgia 30303, USA}
\newcommand{\hiroshima}{Hiroshima University, Kagamiyama, Higashi-Hiroshima 739-8526, Japan}
\newcommand{\ihepprot}{IHEP Protvino, State Research Center of Russian Federation, Institute for High Energy Physics, Protvino, 142281, Russia}
\newcommand{\illuiuc}{University of Illinois at Urbana-Champaign, Urbana, Illinois 61801, USA}
\newcommand{\inrras}{Institute for Nuclear Research of the Russian Academy of Sciences, prospekt 60-letiya Oktyabrya 7a, Moscow 117312, Russia}
\newcommand{\instpasczech}{Institute of Physics, Academy of Sciences of the Czech Republic, Na Slovance 2, 182 21 Prague 8, Czech Republic}
\newcommand{\isu}{Iowa State University, Ames, Iowa 50011, USA}
\newcommand{\jaea}{Advanced Science Research Center, Japan Atomic Energy Agency, 2-4 Shirakata Shirane, Tokai-mura, Naka-gun, Ibaraki-ken 319-1195, Japan}
\newcommand{\jyvaskyla}{Helsinki Institute of Physics and University of Jyv{\"a}skyl{\"a}, P.O.Box 35, FI-40014 Jyv{\"a}skyl{\"a}, Finland}
\newcommand{\kek}{KEK, High Energy Accelerator Research Organization, Tsukuba, Ibaraki 305-0801, Japan}
\newcommand{\korea}{Korea University, Seoul, 136-701, Korea}
\newcommand{\kurchatov}{Russian Research Center ``Kurchatov Institute", Moscow, 123098 Russia}
\newcommand{\kyoto}{Kyoto University, Kyoto 606-8502, Japan}
\newcommand{\labllr}{Laboratoire Leprince-Ringuet, Ecole Polytechnique, CNRS-IN2P3, Route de Saclay, F-91128, Palaiseau, France}
\newcommand{\lahorelums}{Physics Department, Lahore University of Management Sciences, Lahore, Pakistan}
\newcommand{\lawllnl}{Lawrence Livermore National Laboratory, Livermore, California 94550, USA}
\newcommand{\losalamos}{Los Alamos National Laboratory, Los Alamos, New Mexico 87545, USA}
\newcommand{\lpc}{LPC, Universit{\'e} Blaise Pascal, CNRS-IN2P3, Clermont-Fd, 63177 Aubiere Cedex, France}
\newcommand{\lund}{Department of Physics, Lund University, Box 118, SE-221 00 Lund, Sweden}
\newcommand{\maryland}{University of Maryland, College Park, Maryland 20742, USA}
\newcommand{\mass}{Department of Physics, University of Massachusetts, Amherst, Massachusetts 01003-9337, USA }
\newcommand{\michigan}{Department of Physics, University of Michigan, Ann Arbor, Michigan 48109-1040, USA}
\newcommand{\muenster}{Institut fur Kernphysik, University of Muenster, D-48149 Muenster, Germany}
\newcommand{\muhlenberg}{Muhlenberg College, Allentown, Pennsylvania 18104-5586, USA}
\newcommand{\myongji}{Myongji University, Yongin, Kyonggido 449-728, Korea}
\newcommand{\nagasaki}{Nagasaki Institute of Applied Science, Nagasaki-shi, Nagasaki 851-0193, Japan}
\newcommand{\newmex}{University of New Mexico, Albuquerque, New Mexico 87131, USA }
\newcommand{\nmsu}{New Mexico State University, Las Cruces, New Mexico 88003, USA}
\newcommand{\ohio}{Department of Physics and Astronomy, Ohio University, Athens, Ohio 45701, USA}
\newcommand{\ornl}{Oak Ridge National Laboratory, Oak Ridge, Tennessee 37831, USA}
\newcommand{\orsay}{IPN-Orsay, Universite Paris Sud, CNRS-IN2P3, BP1, F-91406, Orsay, France}
\newcommand{\peking}{Peking University, Beijing 100871, P.~R.~China}
\newcommand{\pnpi}{PNPI, Petersburg Nuclear Physics Institute, Gatchina, Leningrad region, 188300, Russia}
\newcommand{\riken}{RIKEN Nishina Center for Accelerator-Based Science, Wako, Saitama 351-0198, Japan}
\newcommand{\rikjrbrc}{RIKEN BNL Research Center, Brookhaven National Laboratory, Upton, New York 11973-5000, USA}
\newcommand{\rikkyo}{Physics Department, Rikkyo University, 3-34-1 Nishi-Ikebukuro, Toshima, Tokyo 171-8501, Japan}
\newcommand{\saopaulo}{Universidade de S{\~a}o Paulo, Instituto de F\'{\i}sica, Caixa Postal 66318, S{\~a}o Paulo CEP05315-970, Brazil}
\newcommand{\stonybrkc}{Chemistry Department, Stony Brook University, SUNY, Stony Brook, New York 11794-3400, USA}
\newcommand{\stonycrkp}{Department of Physics and Astronomy, Stony Brook University, SUNY, Stony Brook, New York 11794-3400, USA}
\newcommand{\tenn}{University of Tennessee, Knoxville, Tennessee 37996, USA}
\newcommand{\titech}{Department of Physics, Tokyo Institute of Technology, Oh-okayama, Meguro, Tokyo 152-8551, Japan}
\newcommand{\tsukuba}{Institute of Physics, University of Tsukuba, Tsukuba, Ibaraki 305, Japan}
\newcommand{\vandy}{Vanderbilt University, Nashville, Tennessee 37235, USA}
\newcommand{\waseda}{Waseda University, Advanced Research Institute for Science and Engineering, 17 Kikui-cho, Shinjuku-ku, Tokyo 162-0044, Japan}
\newcommand{\weizmann}{Weizmann Institute, Rehovot 76100, Israel}
\newcommand{\wigner}{Institute for Particle and Nuclear Physics, Wigner Research Centre for Physics, Hungarian Academy of Sciences (Wigner RCP, RMKI) H-1525 Budapest 114, POBox 49, Budapest, Hungary}
\newcommand{\yonsei}{Yonsei University, IPAP, Seoul 120-749, Korea}
\affiliation{\abilene}
\affiliation{\augie}
\affiliation{\banaras}
\affiliation{\barc}
\affiliation{\baruch}
\affiliation{\bnlcoll}
\affiliation{\bnlphys}
\affiliation{\caucr}
\affiliation{\charlesczech}
\affiliation{\chonbuk}
\affiliation{\ciae}
\affiliation{\cns}
\affiliation{\colorado}
\affiliation{\columbia}
\affiliation{\czechtech}
\affiliation{\dapnia}
\affiliation{\elte}
\affiliation{\ewha}
\affiliation{\fit}
\affiliation{\fsu}
\affiliation{\gsu}
\affiliation{\hiroshima}
\affiliation{\ihepprot}
\affiliation{\illuiuc}
\affiliation{\inrras}
\affiliation{\instpasczech}
\affiliation{\isu}
\affiliation{\jaea}
\affiliation{\jyvaskyla}
\affiliation{\kek}
\affiliation{\korea}
\affiliation{\kurchatov}
\affiliation{\kyoto}
\affiliation{\labllr}
\affiliation{\lahorelums}
\affiliation{\lawllnl}
\affiliation{\losalamos}
\affiliation{\lpc}
\affiliation{\lund}
\affiliation{\maryland}
\affiliation{\mass}
\affiliation{\michigan}
\affiliation{\muenster}
\affiliation{\muhlenberg}
\affiliation{\myongji}
\affiliation{\nagasaki}
\affiliation{\newmex}
\affiliation{\nmsu}
\affiliation{\ohio}
\affiliation{\ornl}
\affiliation{\orsay}
\affiliation{\peking}
\affiliation{\pnpi}
\affiliation{\riken}
\affiliation{\rikjrbrc}
\affiliation{\rikkyo}
\affiliation{\saopaulo}
\affiliation{\stonybrkc}
\affiliation{\stonycrkp}
\affiliation{\tenn}
\affiliation{\titech}
\affiliation{\tsukuba}
\affiliation{\vandy}
\affiliation{\waseda}
\affiliation{\weizmann}
\affiliation{\wigner}
\affiliation{\yonsei}
\author{A.~Adare} \affiliation{\colorado}
\author{C.~Aidala} \affiliation{\mass} \affiliation{\michigan}
\author{N.N.~Ajitanand} \affiliation{\stonybrkc}
\author{Y.~Akiba} \affiliation{\riken} \affiliation{\rikjrbrc}
\author{H.~Al-Bataineh} \affiliation{\nmsu}
\author{J.~Alexander} \affiliation{\stonybrkc}
\author{A.~Angerami} \affiliation{\columbia}
\author{K.~Aoki} \affiliation{\kyoto} \affiliation{\riken}
\author{N.~Apadula} \affiliation{\stonycrkp}
\author{Y.~Aramaki} \affiliation{\cns} \affiliation{\riken}
% ...
\collaboration{PHENIX Collaboration} \noaffiliation

\pacs{25.75.Dw} 
	
\maketitle

\section{Introduction}

Proton- and deuteron-nucleus collisions provide an excellent tool for 
studying a variety of nuclear effects.  For example, there are important 
modifications to parton distribution functions (PDFs) in nuclei, which are 
strongly dependent on $x$ and $Q^2$~\cite{Eskola:2009uj,Helenius:2012wd}; 
the low $x$ partons have a wavelength longer than the extent of large 
nuclei and thus saturation effects are expected to scale with the nuclear 
thickness. Modifications of particle yields due to color neutralization and 
hadronization within the target nucleus are elucidated by their path length 
dependence through the target~\cite{Accardi:2012hwp}. Furthermore, studies 
in deep inelastic scattering (DIS) average over the geometry of a given 
nuclear target, and studies of the geometric dependence of nuclear 
modifications are restricted to varying the nuclear target atomic number 
$A$.  In proton- or deuteron-nucleus collisions, there has been a 
significant effort to characterize the geometry in individual collisions in 
terms of the impact parameter or the number of binary collisions (\Ncoll). 
Being able to characterize the geometry of the collision allows for the 
comparison of \pp and \pda yields through physics quantities, such as the 
nuclear modification factor \rpda

\begin{equation}
\rpda = \frac{dN^{p(d)+A}/dy}{\mNcoll dN^{pp}/dy},
\label{eq:RdA}
\end{equation}
where $dN^{p(d)+A}/dy$ and $dN^{pp}/dy$ are the invariant yields in \pda 
and \pp collisions respectively. Therefore, being able to precisely 
determine the geometric properties of the collision is critical to 
understanding these nuclear effects.

Recent data taken with \ppb collisions at the Large Hadron Collider (LHC) 
and new analysis of \dau collision data at Relativistic Heavy Ion Collider 
(RHIC) indicate possible 
collective flow effects with an important geometric 
dependence~\cite{Adare:2013piz,Aad:2012gla,Abelev:2012ola,CMS:2012qk}.  
These measurements further highlight the need to characterize the geometry 
in individual collisions and for future measurements with different 
collision species.

At lower energies, ``gray tracks'' (named for their appearance in emulsion 
experiments) from spectator nucleons have been utilized for characterizing 
geometry~\cite{Sikler:2003ef,Brenner:1981kf,Chemakin:1999jd}.  At RHIC and 
the LHC, the categorization has been done with backward rapidity 
multiplicity, and not with spectator nucleons.  This general method has 
been utilized by all RHIC experiments in numerous observables over the last 
decade.  In this paper we describe in detail a method for characterizing 
\pda collisions.

The geometry correlation with backward rapidity multiplicity is biased when 
an additional condition on the event is included, for example the 
production of a midrapidity particle.  We describe a procedure for 
correcting the \pda centrality dependent particle yields for this 
auto-correlation bias.  Because high statistics \pda data are available with 
yields extending to high transverse momentum (\pt), it is also important to 
study the dependence of these bias correction factors on the $p_T$ of the 
produced particle.  We utilize measurements in \pp collisions to study the 
correlation of backward rapidity multiplicity with the presence of a high 
\pt midrapidity particle. We discuss this method in the context of the 
PHENIX experiment, though this method is applicable in any collider 
experiment with a far-backward particle detector. In addition, we test this 
method utilizing the {\sc hijing} Monte-Carlo (MC) generator for both \dau 
collisions at \sqsn=~200~GeV and \ppb collisions at \sqsn=~5.02~TeV.

The paper is organized as follows: 
Section~\ref{sec:experiment} discusses the detectors used by the PHENIX 
experiment to characterize the geometry of the collision, which gives 
context to the specific tests detailed later in the paper.  
Sections~\ref{sec:method}-\ref{sec:biasfactors} describe the general 
methodology for centrality categorization.  Section~\ref{sec:method} 
describes the method, Section~\ref{sec:neutroncrosscheck} details 
the cross checks with neutron-tagged events, and Sections~\ref{sec:geometry} 
and~\ref{sec:biasfactors} provide the derivation and systematic 
uncertainties of various geometric quantities, as well as the bias-factor 
corrections to the measured yields.  Section~\ref{sec:hijing} details the 
calculation of the bias factors using the {\sc hijing} MC generator and 
comparisons to those obtained in previous sections with our 
Glauber+negative-binomial-distribution (Glauber+NBD) procedure. 
Section~\ref{sec:summary} summarizes the findings.

\section{Experiment}
\label{sec:experiment}

To characterize the geometry of \dau collisions at \sqsn=~200~GeV, the 
PHENIX experiment uses beam-beam counters (BBCs) covering the 
pseudorapidity range \bbceta and zero-degree calorimeters (ZDCs) covering
$|\eta| > 6$. Both detectors are described in detail in 
Refs.~\cite{Allen:2003zt,Adler:2003sp}.

Each BBC is an array of 64 \v{C}erenkov counters around the beam pipe and 
is positioned 1.44 meters upstream and downstream of the nominal vertex 
location.  Each counter is composed of 3~cm of quartz coupled to a 
mesh-dynode photomultiplier tube.  Although the BBC charge is calibrated 
to a minimum-ionizing charged particle~\cite{Allen:2003zt}, approximately 
50\% of the hits are the result of scattering particles from outside the 
nominal pseudorapidity acceptance of the BBC or of photon conversions 
(e.g., in the Beryllium beam pipe). The information from the BBC is used 
to determine the event timing, vertex position, and centrality.

The two ZDCs are hadronic calorimeters that measure spectator neutrons. 
They are located 18 meters from the interaction point and comprise optical 
readout fibers between tungsten plates. At the top RHIC energy of 
100~GeV/nucleon, neutrons evaporated from the spectator remnants of the 
collision are emitted within 1 mrad from the colliding-beam direction. 
Charged fragments and the noninteracted primary beam are bent by deflecting 
magnets to much larger angles. The ZDC thus measures the total neutron 
energy within a small cone and thus provides the number of spectator 
neutrons from the interacting nucleus.

\section{Centrality Categorization Method}
\label{sec:method}

\begin{figure}[thb]
\includegraphics[width=1.0\linewidth]{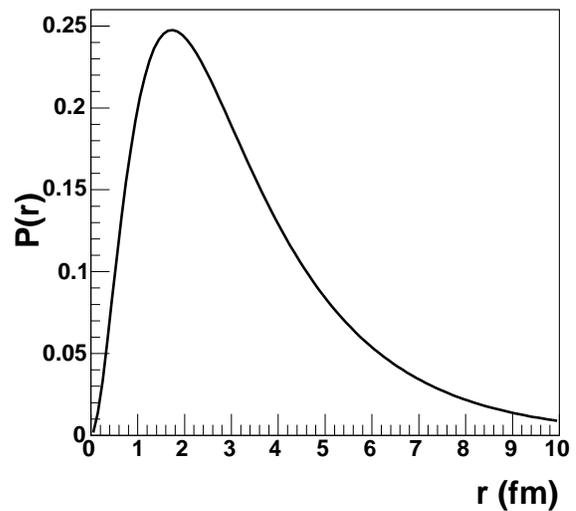}
\caption{\label{fig:hulthen_prob}Probability density distribution for the proton-neutron distance in the deuteron given
  by the square of the Hulth\'en wave function~\cite{hulthen}.}
\end{figure}

\begin{figure}[thb]
\includegraphics[width=1.0\linewidth]{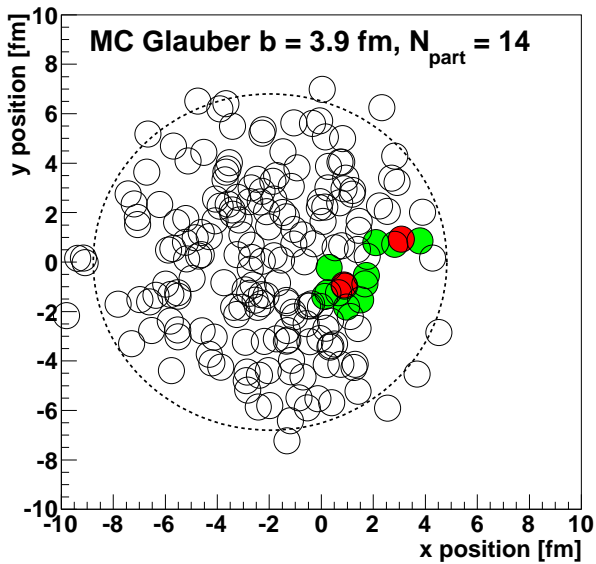}  
\caption{\label{fig:glaubereventdisplay} 
MC-Glauber event display for a single \dau collision.  All 
nucleons are shown as open circles and nucleons with at least one inelastic 
collision are highlighted as filled circles.}
\end{figure}

The PHENIX experiment selects centrality categories in \dau collisions 
based on the summed charge measured in the BBC in the Au-going direction.  
To determine the mapping from the measured charge to various 
geometric quantities such as the number of binary collisions, we employ a 
standard MC-Glauber model.  The general procedure for such 
calculations in heavy ions is described in Ref.~\cite{Miller:2007ri}.  On 
an event-by-event basis, the transverse position of all nucleons in the 
deuteron and gold nucleus are selected from the Hulth\'en wave function and 
a Woods-Saxon distribution, respectively.  The deuteron is described by a 
Hulth\'en wave function:
\begin{equation}
\psi_{d}(r_{pn})=\left( {{\alpha \beta (\alpha+\beta)} \over {2 \pi (\alpha-\beta)^{2}}} \right) ^{1/2} {{(e^{-\alpha r_{pn}}-e^{-\beta r_{pn}})} \over {r_{pn}}},
\end{equation}
with $\alpha$ = 0.228 fm$^{-1}$ and $\beta$ = 1.18 
fm$^{-1}$~\cite{hulthen}.  The square of this wave function determines the 
probability distribution for the distance between the proton and neutron 
within the deuteron, as shown in Fig.~\ref{fig:hulthen_prob}.  For the 
gold nucleus, we use the Woods-Saxon density distribution:
\begin{equation}
\rho(r) = \frac{\rho_0}{1+e^{\frac{r-R}{a}}},
\label{eq:WS}
\end{equation}
with radius $R=6.38$ fm and diffuseness parameter $a=0.54$ fm. We utilize a 
nucleon-nucleon inelastic cross section of 42 mb.  Variation of these 
values and the inclusion of a hard-core repulsive potential are utilized in 
the determination of systematic uncertainties described later.  On an 
event-by-event basis, the nucleon positions are determined at random 
(weighted with the respective probability distributions), an impact 
parameter is selected, and the nucleon-nucleon collisions are calculated. A 
nucleon-nucleon collision occurs if the distance between two nucleons is 
less than $\sqrt{\sigma_{NN}/\pi}$.  A single \dau event is shown in 
Fig.~\ref{fig:glaubereventdisplay}, where both nucleons from the deuteron, 
shown as red circles, have inelastic collisions and the nucleons from the 
gold nucleus, shown as green circles, with at least one inelastic collision 
are highlighted.  Once the participating nucleons are determined in a given 
event, one has full information on the number of participating nucleons, 
the number of binary collisions, and the spatial position and overlap of 
all participating nucleons.

\begin{figure}[thb]
\includegraphics[width=1.0\linewidth]{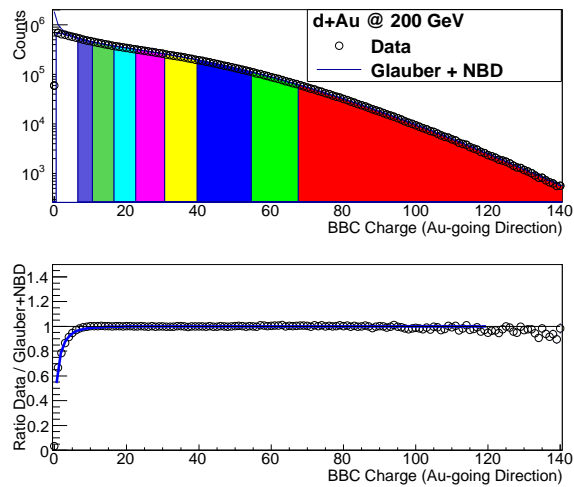} 
\caption{\label{fig:bbccharge} (upper) Real data \dau BBC Au-going 
direction charge distribution are shown as open points and Glauber+NBD 
calculation as a histogram.  Shown as different color regions are the 
centrality selections of 0\%--5\%, 5-10\%, 10\%--20\%, 20\%--30\%, 
30\%--40\%, 40\%--50\%, 50\%--60\%, 60\%--70\%, and 70\%--88\%.  (lower) 
The ratio of real data to Glauber+NBD calculation.  The line is a fit to 
the experimental trigger efficiency turn-on curve.  }
\end{figure}

To relate these MC-Glauber quantities to geometric 
parameters, they must be mapped to an experimental observable.  The PHENIX 
experiment has used as the experimental observable the charge measured in 
the BBC in the Au-going direction covering pseudorapidity -3.9 $< \eta <$ 
-3.0.  The minimum-bias (MB) trigger requirement is the coincidence of one 
or more hits in both the BBC in the Au-going direction and in the BBC in 
the $d$-going direction. The experimental distribution of the BBC Au-going 
charge corresponding to this MB trigger sample is shown as open 
circles in the upper panel of Fig.~\ref{fig:bbccharge}.

At this point we make the hypothesis that the BBC Au-going charge is 
proportional to the number of binary collisions in an individual \dau 
collision, with fluctuations in the contribution from each binary collision 
described by the NBD, which is parametrized by the mean $\mu$ and a 
positive exponent $\kappa$:
\begin{equation}
\label{eqn:nbd}
{\rm NBD}(x;\mu ,\kappa) = \left(1+\frac{\mu}{\kappa} \right ) 
\frac{(\kappa+x-1)!}{x!(\kappa-1)!}\left(\frac{\mu}{\mu+\kappa}\right)^x.
\end{equation}
NBD distributions have been utilized for fitting particle multiplicities, 
see for example Refs.~\cite{Giovannini:1985mz,PhysRevC.52.2663}, in part 
due to the fact that randomly sampling from $n$ NBD($\mu,\kappa$) 
distributions results in an NBD distribution with NBD($n\mu,n\kappa$).  
The fluctuations contained in the NBD relate both to the variation in the 
number and distribution of particles produced, and also to fluctuations in 
the number of particles resulting in charge in the BBC detector.  One then 
folds the Glauber distribution of the number of binary collisions 
($Gl(n)$), normalized per event, with the NBD response using
\begin{equation}
P(x) = \sum_{n=1}^{N_{binary}(max)} Gl(n)\times{\rm NBD}(x;n\mu,n\kappa),
\end{equation}
where $x$ is the BBC charge.  

The two NBD parameters $\mu$ and $\kappa$ are fit to the experimental 
distribution for BBC charge greater than 20.  At low BBC charge there will 
be an expected deviation between the calculation and data due to the 
inefficiency of the MB trigger requirement (including at least 
one hit in the $d$-going BBC).  The result of the best fit yields values 
$\mu = 3.03$ and $\kappa = 0.46$ and is shown as a histogram in the upper 
panel of Fig.~\ref{fig:bbccharge}. The ratio of the data to the 
Glauber+NBD calculation is shown in the lower panel of 
Fig.~\ref{fig:bbccharge} and shows very good agreement for BBC charge 
greater than 10. We have also considered the possibility that the Au-going 
charge is proportional to the number of Au participants, rather than the 
number of binary collisions. We get an equally good fit to the data, and 
the difference in the extracted geometric quantities (detailed later) is 
included in the quoted systematic uncertainty.

We have fit the deviation at low charge to determine the MB 
trigger efficiency turn on curve.  The integrated results indicate that the 
MB trigger fires on 88 $\pm$ 4\% of the Glauber determined 2.19 
barn inelastic \dau cross section.  The PHENIX experiment has separately 
measured the MB trigger sample cross section.  The deuteron 
dissociation cross section $\sigma(d \rightarrow p+n)$ is theoretically 
well calculated as 1.38 $\pm$ 0.01 barns and thus combining this with the 
measured ratio of MB to dissociation cross section, one obtains 
the MB cross section of 1.99$\pm$0.10 barns~\cite{White:2005kp}.  
When combined with the 88\% trigger efficiency this yields a total 
inelastic cross section of 2.26 $\pm$ 0.10 barns, in agreement with the 
previous value.

\begin{figure}[thb]
\includegraphics[width=1.0\linewidth]{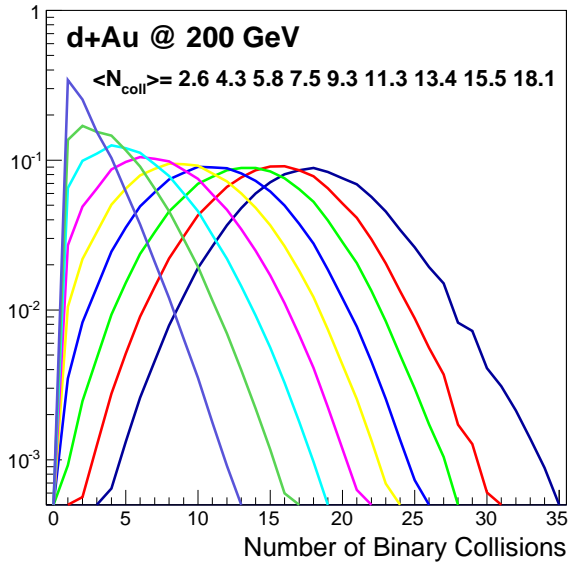}  
\caption{\label{fig:ncolldist} Extracted distribution of the number of 
binary collisions in each of the nine centrality quantiles: 0\%--5\%, 
5-10\%, 10\%--20\%, 20\%--30\%, 30\%--40\%, 40\%--50\%, 50\%--60\%, 
60\%--70\%, and 70\%--88\%.  }
\end{figure}

\begin{figure}[thb]
\includegraphics[width=1.0\linewidth]{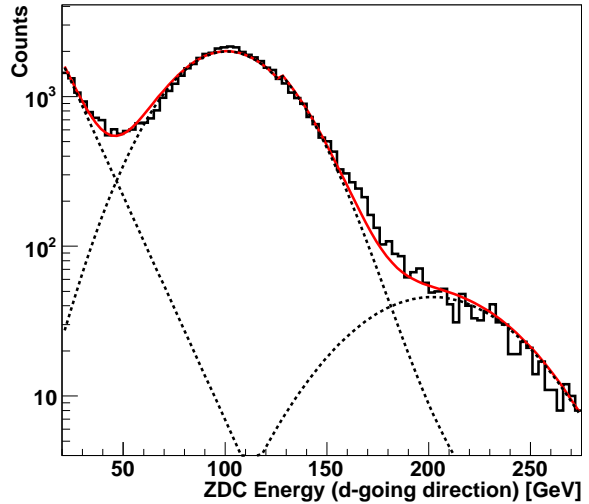}  
\caption{\label{fig:zdcenergy} 
ZDC energy distribution in the deuteron-going direction for MB 
\dau collisions. The data are well described by an exponential background 
component, a single spectator neutron peak, and a much smaller contribution 
from two neutrons due to double interactions.}
\end{figure}

In \pp collisions there is always one binary collision.  The NBD parameters 
determined above are consistent within uncertainties with the mean BBC 
multiplicity in \pp collisions.  However, an exact comparison of the full 
distribution is challenging, because the MB trigger significantly 
biases the distribution. Utilizing Clock Triggers (random triggers with no 
detector requirement) has the difficulty of background contamination from 
beam-gas interactions and beam-beam collisions outside the nominal PHENIX 
z-vertex range ($-30~{\rm cm} < z < +30$~cm).

For each centrality bin we can apply the identical event selection and 
trigger turn-on curve on the Glauber+NBD calculation, thus determining the 
distribution of the number of binary collisions (number of nucleon-nucleon 
inelastic collisions).  The results corresponding to the nine centrality 
quantiles 0\%--5\%, 5-10\%, 10\%--20\%, 20\%--30\%, 30\%--40\%, 40\%--50\%, 
50\%--60\%, 60\%--70\%, 70\%--88\% are shown in Fig.~\ref{fig:ncolldist}.

\section{Neutron-Tagged Cross Check}
\label{sec:neutroncrosscheck}

As an additional test, we checked the validity of our geometry selection 
method using neutron-tagged events.  Due to the large size of the deuteron, 
there is a significant probability for the neutron (proton) from the 
deuteron to miss the Au nucleus (i.e. have no inelastic interaction with 
any target nucleon) while the proton (neutron) does interact.  These 
``$p$''+Au and ``$n$''+Au interactions have been studied and are detailed 
in Ref.~\cite{Adler:2007aa}.  In the ``$p$''+Au case, the method employed 
is to measure the spectator neutron energy in the PHENIX Zero Degree 
Calorimeter (ZDC) in the deuteron-going direction.  The ZDC energy 
distribution in \dau MB events is shown in 
Fig.~\ref{fig:zdcenergy}. The distribution is only for events where 
energy above threshold is deposited in the ZDC, and therefore the majority 
of events, i.e. those where there is no spectator neutron, are not 
included. One observes a clear single neutron peak with a mean energy of 
100~GeV (the expected beam energy) and a resolution width of approximately 
28~GeV.  Additionally, there is a low energy background component that is 
well described by an exponential.  Lastly there is a contribution from two 
neutrons.  The two neutron contribution comes from double interactions 
where the additional neutron results from an independent inelastic \dau 
interaction or a \dau photo-disintegration reaction.

\begin{figure}[thb]
\includegraphics[width=1.0\linewidth]{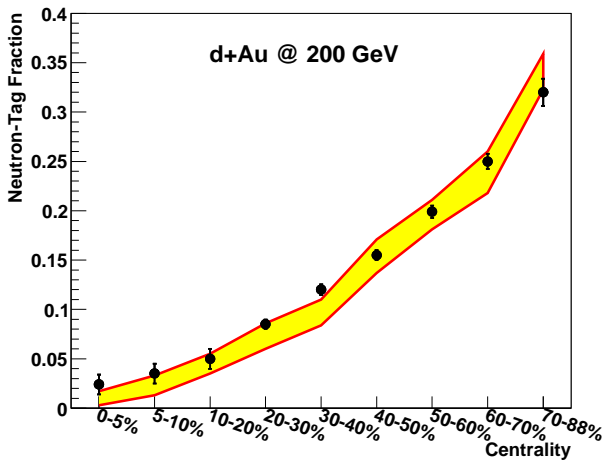}  
\caption{\label{fig:ntagfrac} 
Data points are the measured fraction of events where there is a spectator 
neutron from the deuteron projectile.  In comparison, the yellow band is 
the MC-Glauber result with systematic uncertainties.  }
\end{figure}

We select events with a spectator neutron with a ZDC energy cut of 
60--180~GeV, which captures 96\% of the single neutron peak.  We estimate 
a 2\%-3\% contribution from the exponential background.  These effects 
tend to cancel and we apply no net correction to the spectator neutron 
event yield, and apply a $\pm3$\% systematic uncertainty on this yield. 
The double interaction contribution (i.e. the two-neutron peak yield) 
depends on the instantaneous luminosity and the ``centrality'' category 
of selected \dau events. Accounting for these double interaction 
contributions as detailed in the next section, we determine from data the 
probability of a spectator neutron from a single \dau inelastic 
interaction in the nine centrality selections, as shown in 
Fig.~\ref{fig:ntagfrac}.  The error bars reflect systematic uncertainties 
from accounting for double interaction contributions between the 
different data sets (dominant in central events) and from the neutron 
tagging efficiency (dominant in peripheral events).  The yellow band 
corresponds to the MC-Glauber calculated values and the systematic 
uncertainties in that calculation from a full set of parameter 
variations, discussed in detail in the next section.  The agreement 
between data and calculation is good and gives us confidence in the 
geometric modeling of the collisions.

\begin{figure}[thb]
\includegraphics[width=0.96\linewidth]{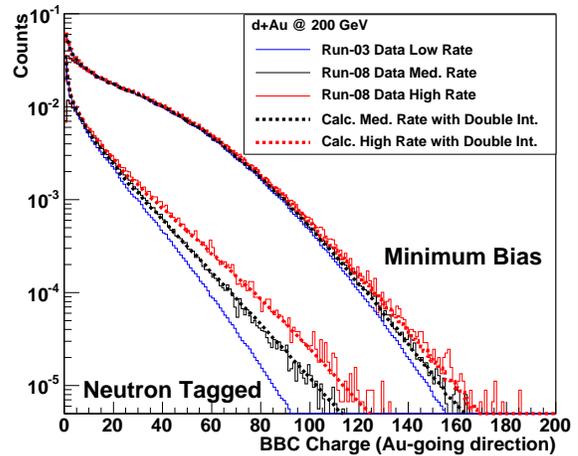}  
\caption{\label{fig:doubleint} 
Data distributions for the BBC Charge (in the Au-going direction) from low 
luminosity data recorded in 2003~\protect\cite{PhysRevC.77.014905}, 
medium-luminosity data from early in 2008, and high-luminosity data from 
late in 2008.  The upper set of curves are for MB \dau interactions 
and the lower set of curves are with the additional requirement of a 
neutron spectator from the deuteron.  The differences in the distributions 
are the result of contributions from double interactions.  The results 
of a calculation of these contributions with input rates relevant to 
match the medium- and high-luminosity 2008 data are also shown.
}
\end{figure}

\section{Double Interaction Study}
\label{sec:doubleinteraction}

Figure~\ref{fig:doubleint} shows the BBC Au-going charge distribution in 
MB \dau events (upper curves) and the distribution in the subset 
of events where there is a single neutron spectator present (lower curves).  
The lowest curve (blue) in each set is from low-luminosity \dau data collected 
in 2003~\cite{PhysRevC.77.014905}.  The middle curve (black) and upper curve (red) 
are from medium and high luminosity data collected in the 2008 \dau 
running, respectively.  The highest luminosities achieved correspond to BBC 
trigger rates of order 300 kHz, where the probability of a second inelastic 
interaction within the same crossing approaches a few percent.  Although 
the probability is low, the case of two inelastic interactions occurring in 
the same crossing results in the sum of their respective BBC Au-going 
charge, and thus a higher probability for a falsely identified \dau central 
event.  In the high luminosity category and for the most central 5\% of 
interactions, the double interaction contribution is approximately 13\%.

The difference in BBC distributions is even more pronounced between data 
taken at different luminosities for the neutron spectator sample.  The 
reason is that a central, high BBC multiplicity event has a small 
probability to have a neutron spectator, because both nucleons from 
the deuteron are typically occluded by the Au nucleus.  Thus, the probability of two 
interactions, one with high multiplicity and no spectator neutron and one 
low multiplicity with a spectator neutron, dominates.  In addition, the 
second interaction may be a photo-dissociation reaction that results in no 
BBC multiplicity contribution and just the neutron striking the ZDC.  
Using a Weizsacker-Williams approach, the photo-dissociation \dau cross 
section is calculated to be 1.38 barns~\cite{PhysRevC.68.017902}.

We employ a simple model to account for all these contributions.  The 2003 
distributions shown in Fig.~\ref{fig:doubleint} are from very-low-luminosity 
data~\cite{PhysRevC.77.014905}, where double interactions are negligible.  
Thus, we use these distributions as from strictly single interactions.  For 
the 2008 medium rate and high rate data samples, given an input instantaneous 
luminosity, we calculate the probability of two inelastic collisions and 
their resulting summed BBC charge.  We also calculate the probability of 
one inelastic collision with no spectator neutron and a second interaction 
(inelastic or photo-disintegration) with a resulting neutron.  We then sum 
the different relative single and double interaction contributions.  Using 
a double interaction probability of 1.5\% (3.5\%), in agreement with the 
instantaneous luminosity calculation for the medium (high) rate period, we 
obtain the thick black (red) dashed lines, which simultaneously match the 
MB and neutron-tagged real-data samples.

The BBC Au-going charge distribution shown in Fig.~\ref{fig:bbccharge} is 
from a low luminosity run in 2008.  The centrality quantiles are determined 
on a run-by-run basis and thus the highest luminosity runs will have some 
influence from the double interaction contamination. In the central 
0\%--20\% sample the contamination is very small for both the 2003 and 2008 
luminosities. Only in the most central 0\%--5\% sample, where the 
contamination can reach as high as 13\% at the highest luminosities, does 
this become a real concern and further checks become necessary. For 
example, an analysis of low-\pt  correlations~\cite{Adare:2013piz} using 
the 0\%--5\% sample, checked explicitly the robustness of the result using  
subsets of low and high luminosity and found good agreement between the two.

\section{Geometry Characterization and Systematic Uncertainties}
\label{sec:geometry}

At this point, with the mapping of BBC charge to MC-Glauber, we 
determine several geometric properties from the MB \dau sample, 
such as \Npart, \Ncoll, and eccentricity, in each centrality selection.  We 
note that an additional requirement of a particular particle at midrapidity 
will bias this geometric mapping due to auto-correlations with the backward 
rapidity multiplicity.  We correct for this effect separately as discussed 
in the next section. To determine the systematic uncertainty on 
these geometric properties, we vary the input parameters and re-run the 
entire NBD parameter fit and trigger efficiency 
turn-on curve determination.  The following variations are considered: 

\begin{enumerate} 

\item We vary the nucleon-nucleon inelastic cross section from the default 
value of 42 mb down to 39 mb and up to 45 mb.

\item We vary the Woods-Saxon Au nucleus parameters. The alternate set \#1 
has a radius = 6.65 fm and diffusiveness = 0.55 fm. The alternate set \#2 
has a radius = 6.25 fm and diffusiveness = 0.53 fm. These are compared to 
the default values of radius = 6.38 fm and diffusiveness = 0.54 fm.

\item We include a hard-core repulsive potential with an exclusion radius 
$r=0.4$ fm in the Glauber model selection of nucleon positions in the Au 
nucleus.  The hard-core potential prevents nucleons from occupying the same 
space.

\item We consider the possibility that the BBC Au-going direction charge is 
not simply proportional to the number of binary collisions, but instead is 
proportional to the number of binary collisions to the power $\alpha$. We 
consider values of $\alpha$ = 0.95 and $\alpha$ = 1.05.  These values of 
$\alpha$ result in a change in particle production per binary collision 
from peripheral events to central events of order 10\%--15\% and are at the 
extreme of consistency with the centrality dependence of the charge 
multiplicity.

\item We run the real data to Glauber+NBD comparison by default for the 
z-vertex range $|z|<5$~cm.  We include comparisons with the extremes in 
the z-vertex selection of 25 to 30~cm and -25 to -30~cm.  The BBC 
acceptance varies slightly with the change in the collision z-vertex and 
due to the centrality-dependent asymmetric rapidity distribution of charged 
particles~\cite{PhysRevC.72.031901}, the variation is different for extreme 
positive and negative vertices.

\end{enumerate}

\begin{figure*}[!thb]
\includegraphics[width=0.99\linewidth]{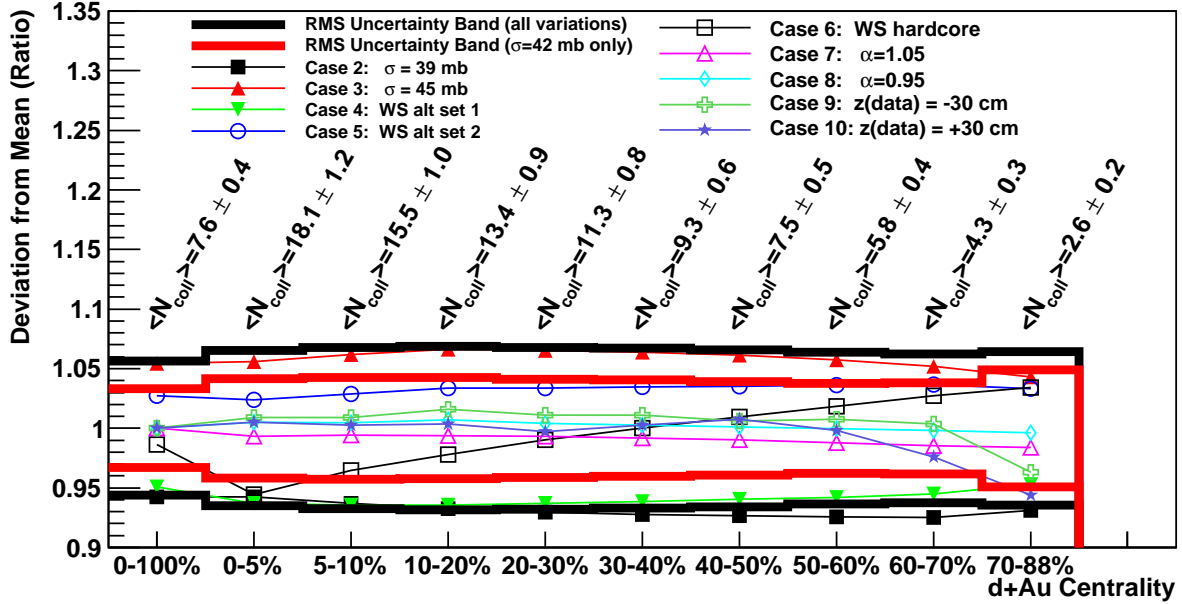}  
\caption{\label{fig:ncollsys} 
The variation relative to the mean \Ncoll value for variations in the input 
assumptions for the MC-Glauber extraction.  The final mean and RMS 
systematic uncertainties are shown in Table~\ref{tab:full9}.
}
\end{figure*}

As an example, the variations in the mean number of binary collisions as 
the inputs are individually changed are shown with different colors in 
Fig.~\ref{fig:ncollsys}.  The largest change in the number of binary 
collisions results when the value for $\sigma_{NN}$ is varied.  In a 
similar manner, changing the Woods-Saxon parameters creates a more (less) 
dense nucleus and moves the \mNcoll values for all centralities up (down).  
The simplistic inclusion of a hardcore repulsive potential between nucleons 
results in the largest absolute shift from central to peripheral as it 
moves nucleons away from the core of the nucleus.  Changes in the scaling 
of the multiplicity ($\alpha$) and the z-vertex range fit result in very 
modest changes over all centralities.

We then consider 81 different scenarios with combinations of these 
variations and re-run the entire procedure.  The mean and the 
root-mean-square (RMS) of these 81 variations yield the final quoted value 
and systematic uncertainty.  These values are also shown in 
Fig.~\ref{fig:ncollsys}.  The black band is the fractional RMS 
uncertainty.  The red band is the fractional RMS uncertainty but without 
the variation in the nucleon-nucleon cross section of 42 mb.  When 
calculating quantities, such as the nuclear modification factor $R_{dAu}$, 
this uncertainty cancels, because the same value appears in determining the 
\pp cross section in the denominator.  The same procedure of 81 variations 
was applied in calculating the neutron spectator fraction shown earlier in 
Fig.~\ref{fig:ntagfrac}.

We have applied the identical procedure to other interesting geometric 
quantities and calculated the mean value and systematic uncertainty for the 
nine centrality selections, plus the integral over all centralities, 
corresponding to 0\%--100\% of the \dau inelastic cross section.  We quote 
the mean number of binary collisions \mNcoll, the total number of 
participating nucleons $\langle\Npart\rangle$, the number of participating 
nucleons from the Au nucleus $\langle\Npart[{\rm Au}]\rangle$, and the 
number of participating nucleons from the deuteron 
$\langle\Npart[d]\rangle$.  We also know the spatial distribution in 
the transverse plane of all the participants, and can thus calculate 
different geometric quantities.  The eccentricity $\varepsilon_{2}$ and 
higher moments are calculated as:
\begin{equation}
\varepsilon_{n} = {{\sqrt{\langle r^{2}cos(n\phi)\rangle^{2}+\langle r^{2}sin(n\phi)\rangle^{2}}} \over {\langle r^{2}\rangle}}
\end{equation}
where $n$ is the $n$th moment of the spatial anisotropy calculated relative 
to the mean position.  The averages are taken over the spatial distribution 
of participating nucleons from the MC-Glauber calculation.  We 
also calculate often used quantities, including the spatial overlap area:
\begin{equation}
S = 4 \pi \sqrt{\langle x^{2}\rangle\langle y^{2}\rangle-\langle xy\rangle^{2}}
\end{equation}
and
\begin{equation}
\overline{R} = {{1} \over {\sqrt{1/\langle x^{2}\rangle+1/\langle y^{2}\rangle}}}
\end{equation}
where these quantities are calculated in a frame rotated to align with the 
second order participant plane.  We have performed these geometric 
calculations assuming 
\begin{enumerate}
\item an equal weighting for each participating nucleon at the point-like 
center position [point], 
\item an equal weighting for each participating nucleon with the spatial 
distribution of a Gaussian with $\sigma = 0.4$ fm [Gauss], 
\item an equal weighting for each participating nucleon with a spatial 
distribution of a uniform disk with $R = 1$ fm [disk], and 
\item with a weighting determined randomly from our NBD calculations  
for each participating nucleon with a spatial distribution 
of a uniform disk with $R = 1$ fm [disk-nbd].  
\end{enumerate}
Note that the 
systematic uncertainty quoted on each quantity includes the 81 variations 
in our standard calculation, and does not include any uncertainty related 
to the assumption used in each of the given four cases above. All of these 
results are given in Table~\ref{tab:full9}.

\begingroup \squeezetable
\begin{table*} 
\caption{Different physical quantities characterizing \dau collisions, 
and the bias-factor corrections, for nine PHENIX centrality bins.}
\begin{ruledtabular} \begin{tabular}
{lc@{\hskip 0.1in}c@{\hskip 0.1in}c@{\hskip 0.1in}c@{\hskip 0.1in}c@{\hskip 0.1in}}
& 0\%--100\% & 0\%--5\% & 5\%--10\% & 10\%--20\% & 20\%--30\% \\
\hline
Bias-Factor Correction & $0.889\pm 0.003$ & $0.91\pm 0.01$ & $0.940\pm 0.007$ & $0.966\pm 0.008$ & $0.99\pm 0.01$ \\

\mNcoll                & $7.6\pm 0.4$ & $18.1\pm 1.2$ & $15.5\pm 1.0$ & $13.4\pm  0.9$ & $11.2\pm 0.8$ \\

$\langle \Npart \rangle$ & $8.6\pm 0.4$ & $17.8\pm 1.2$ & $15.6\pm 1.0$ & $14.1\pm 0.8$ & $11.9\pm 0.7$ \\

$\langle \Npart [{\rm Au}] \rangle$ & $7.0\pm 0.4$ & $15.8\pm 1.2$ & $13.6\pm 1.0$ & $12.2\pm 0.8$ & $10.0\pm 0.7$ \\

$\langle\Npart [d] \rangle$ & $1.62\pm 0.01$ & $1.97\pm 0.02$ & $1.96\pm 0.02$ & $1.91\pm 0.02$ & $1.88\pm 0.03$ \\

$\langle\epsilon_2 \rangle$ (pt-like) & $0.70\pm 0.01$ & $0.63\pm 0.04$ & $0.64\pm 0.02$ & $0.64\pm 0.02$ & $0.63\pm 0.02$ \\

$\sqrt{\langle\epsilon_2^2\rangle}$ (pt-like) & $0.75\pm 0.01$ & $0.67\pm 0.04$ & $0.68\pm 0.02$ & $0.68\pm 0.02$ & $0.68\pm 0.02$ \\

$\langle\epsilon_2 \rangle$ (Gauss-like) & $0.453\pm 0.007$ & $0.54\pm 0.04$ & $0.54\pm 0.02$ & $0.53\pm 0.02$ & $0.51\pm 0.02$ \\

$\langle\epsilon_2 \rangle$ (disk-like) & $0.390\pm 0.007$ & $0.49\pm 0.04$ & $0.50\pm 0.02$ & $0.48\pm 0.02$ & $0.47\pm 0.02$ \\

$\langle\epsilon_2 \rangle$ (disk-nbd) & $0.415\pm 0.008$ & $0.50\pm 0.04$ & $0.51\pm 0.02$ & $0.50\pm 0.02$ & $0.49\pm 0.02$ \\

$\langle S\rangle$ (pt-like) & $4.36\pm 0.24$ & $8.0\pm 0.8$ & $7.6\pm 0.4$ & $7.2\pm 0.4$ & $6.8\pm 0.4$ \\

$\langle S\rangle$ (Gauss-like) & $7.04\pm 0.24$ & $10.8\pm 0.8$ & $10.4\pm 0.8$ & $10.0\pm 0.4$ & $9.6\pm 0.4$ \\

$\langle S\rangle$ (disk-like) & $8.56\pm 0.24$ & $12.0\pm 0.8$ & $12.0\pm 0.8$ & $11.6\pm 0.4$ & $11.2\pm 0.4$ \\

$\langle S\rangle$ (disk-nbd) & $6.96\pm 0.2$ & $10.8\pm 0.8$ & $10.4\pm 0.4$ & $10.0\pm 0.4$ & $9.2\pm 0.4$ \\

$\langle \overline{R}\rangle$ (pt-like) & $0.28\pm 0.01$ & $0.45\pm 0.02$ & $0.43\pm 0.02$ & $0.42\pm 0.02$ & $0.40\pm 0.02$ \\

$\langle \overline{R}\rangle$ (Gauss-like) & $0.449\pm 0.008$ & $0.55\pm 0.02$ & $0.55\pm 0.02$ & $0.53\pm 0.01$ & $0.52\pm 0.01$ \\ 

$\langle \overline{R}\rangle$ (disk-like) & $0.513\pm 0.007$ & $0.61\pm 0.02$ & $0.60\pm 0.01$ & $0.59\pm 0.01$ & $0.58\pm 0.01$ \\

$\langle \overline{R}\rangle$ (disk-nbd) & $0.455\pm 0.006$ & $0.57\pm 0.02$ & $0.56\pm 0.01$ & $0.54\pm 0.01$ & $0.53\pm 0.01$ \\

$\langle\epsilon_3 \rangle$ (pt-like) & $0.311\pm 0.004$ & $0.26\pm 0.02$ & $0.28\pm 0.02$ & $0.30\pm 0.01$ & $0.31\pm 0.01$ \\

$\langle\epsilon_3 \rangle$ (Gauss-like) &  $0.174\pm 0.004$ & $0.19\pm 0.01$ & $0.20\pm 0.01$ & $0.21\pm 0.01$ & $0.208\pm 0.008$ \\

\hline
& 30\%--40\% & 40\%--50\% & 50\%--60\% & 60\%--70\% & 70\%--88\% \\
\hline
Bias-Factor Correction& $1.01\pm 0.02$ & $1.03\pm 0.02$ & $1.050\pm 0.003$ & $1.07\pm 0.06$ & $1.1\pm 0.1$\\

\mNcoll &                $9.3\pm 0.6$   & $7.5\pm 0.5$ &    $5.8\pm 0.4$ &    $4.2\pm 0.3$  & $2.6\pm 0.2$\\

$\langle \Npart \rangle$ & $10.5\pm 0.6$ & $8.7\pm 0.5$ & $7.1\pm 0.4$ & $5.7\pm 0.4$ & $3.9 \pm 0.3$\\

$\langle \Npart [{\rm Au}] \rangle$ & $8.7\pm 0.6$ & $7.0\pm0.5$ & $5.5\pm 0.4$ & $4.1\pm 0.3$ & $2.6\pm 0.3$\\

$\langle\Npart [d] \rangle$ & $1.82\pm 0.04$ & $1.70\pm 0.03$ & $1.62\pm 0.03$ & $1.55\pm 0.06$ & $1.30\pm 0.03$\\

$\langle\epsilon_2 \rangle$ (pt-like) &  $0.63\pm 0.02$ & $0.63\pm 0.03$ & $0.66\pm 0.02$ & $0.71\pm 0.02$ & $0.80\pm 0.02$\\ 

$\sqrt{\langle\epsilon_2^2\rangle}$ (pt-like) &  $0.68\pm 0.02$ & $0.67\pm 0.03$ & $0.71\pm 0.02$ & $0.76\pm 0.02$ & $0.84\pm 0.02$\\

$\langle\epsilon_2 \rangle$ (Gauss-like) & $0.49\pm 0.02$ & $0.45\pm 0.03$ & $0.44\pm 0.02$ & $0.43\pm 0.02$ & $0.39\pm 0.01$\\

$\langle\epsilon_2 \rangle$ (disk-like) & $0.44\pm 0.01$ & $0.40\pm 0.03$ & $0.38\pm 0.02$ & $0.36\pm 0.02$ & $0.31\pm 0.01$\\

$\langle\epsilon_2 \rangle$ (disk-nbd) & $0.47\pm 0.02$ & $0.44\pm 0.03$ & $0.43\pm 0.02$ & $0.39\pm 0.02$ & $0.33\pm 0.01$\\ 

$\langle S\rangle$ (pt-like) &  $6.0\pm 0.4$ & $4.8\pm 0.4$ & $4.00\pm 0.36$ & $2.8\pm 0.4$ & $1.64\pm 0.02$\\

$\langle S\rangle$ (Gauss-like) & $8.8\pm 0.4$ & $7.6\pm 0.4$ & $6.64\pm 0.36$ & $5.6\pm 0.4$ & $4.32\pm 0.24$\\          

$\langle S\rangle$ (disk-like) & $10.0\pm 0.4$ & $9.2\pm 0.4$ & $8.12\pm 0.36$ & $7.2\pm 0.4$ & $5.72\pm 0.24$\\

$\langle S\rangle$ (disk-nbd) &  $8.44\pm 0.36$ & $7.2\pm 0.4$ & $6.4\pm 0.4$ & $5.48\pm 0.36$ & $4.28\pm 0.2$\\

$\langle \overline{R}\rangle$ (pt-like) &  $0.37\pm 0.02$ & $0.33\pm 0.02$ & $0.29\pm 0.02$ & $0.23\pm 0.02$ & $0.14\pm 0.01$\\

$\langle \overline{R}\rangle$ (Gauss-like) & $0.50\pm 0.01$ & $0.47\pm 0.01$ & $0.44\pm 0.01$ & $0.41\pm 0.01$ & $0.370\pm 0.007$\\

$\langle \overline{R}\rangle$ (disk-like) &  $0.56\pm 0.01$ & $0.53\pm 0.01$ & $0.50\pm 0.01$ & $0.48\pm 0.01$ & $0.440\pm 0.006$\\

$\langle \overline{R}\rangle$ (disk-nbd) & $0.50\pm 0.01$ & $0.48\pm 0.01$ & $0.44\pm 0.01$ & $0.42\pm 0.01$ & $0.371\pm 0.007$\\

$\langle\epsilon_3 \rangle$ (pt-like) & $0.35\pm 0.01$ & $0.37\pm 0.02$ & $0.38\pm 0.02$ & $0.36\pm 0.02$ & $0.29\pm 0.02$\\

$\langle\epsilon_3 \rangle$ (Gauss-like) &  $0.214\pm 0.009$ & $0.21\pm 0.01$ & $0.193\pm 0.008$ & $0.17\pm 0.01$ & $0.129\pm 0.009$\\

\end{tabular} \end{ruledtabular}
\label{tab:full9}
\end{table*}
\endgroup

In addition, many PHENIX results have been categorized in four centrality 
selections. For completeness we quote those in Table~\ref{tab:full4}.

\begin{table}
\caption{Different physical quantities characterizing \dau collisions, and 
the bias-factor corrections, for four PHENIX centrality bins.}
\begin{ruledtabular} \begin{tabular}{lcccc}
& 0\%--20\% & 20\%--40\% & 40\%--60\% & 60\%--88\% \\
\hline 
\parbox{1.7cm}{\vspace{1mm} Bias-Factor\\Correction}                     
& $0.94\pm 0.01$ & $1.00\pm 0.01$ & $1.03\pm 0.02$ & $1.03\pm 0.06$ \\
\mNcoll                          & $15.1\pm 1.0$ & $10.2\pm 0.7$ & $6.6\pm 0.4$ & $3.2\pm 0.2$ \\
$\langle\Npart\rangle$           & $15.2\pm 0.6$ & $11.1\pm 0.6$ & $7.8\pm 0.4$ & $4.3\pm 0.2$ \\
$\langle\Npart[{\rm Au}]\rangle$ & $13.3\pm 0.8$ &  $9.3\pm 0.6$ & $6.2\pm 0.4$ & $3.0\pm 0.2$ \\
$\langle\Npart[d]\rangle$        & $1.95\pm 0.01$ & $1.84\pm 0.01$ & $1.65\pm 0.02$ & $1.36\pm 0.02$ \\
\end{tabular} \end{ruledtabular}
\label{tab:full4}
\end{table}

\section{Bias-Factor Corrections}
\label{sec:biasfactors}

At this point we can calculate the invariant yield of a given particle for 
a given centrality selection and correlate that with the geometric 
quantification described above.  In this section to take into account 
the auto-correlations between the presence of a particular particle 
and the backward rapidity multiplicity, we look for additional corrections 
to the previous Glauber+NBD model.  This correction can then be 
multiplied by the invariant yield, of $\pi^0$ at midrapidity (for example), 
to give an accurate match to the quantities derived in the previous 
section.

Here we discuss a specific auto-correlation bias in \pp collisions at 
\sqsn=200~GeV. The PHENIX experiment has measured that inelastic \pp 
reactions, corresponding to the 42 mb cross section, fire the BBC trigger 
$52 \pm 4$\% of the time.  In contrast, a \pp collision with a 
midrapidity-produced pion, charged hadron or $J/\psi$ fires the trigger
~75 $\pm$ 3\% of the time.  The simple reason is that the multiplicity 
in these events is higher.  The \pp 42 mb inelastic cross 
section can be thought of as having three distinct 
contributions~\cite{Adler:2007aa}: 
\begin{enumerate}
\item nondiffractive collisions with 28 mb, 
\item single diffractive collisions with 10 mb, and 
\item double diffractive collisions with 4 mb. 
\end{enumerate}
A {\sc pythia} 6.2 MC simulation of \pp collisions coupled with a 
{\sc geant} modeling of the BBC yields trigger efficiencies of 72 $\pm$ 1\%, 
7 $\pm$ 1\% and 32 $\pm$ 1\%, for each process respectively.  Single and 
double diffractive collisions produce particles dominantly near the beam 
rapidity and thus have a small probability for particle production in the 
BBC acceptance of \bbceta, and an even smaller probability at midrapidity. 
The BBC trigger is therefore biased to the nondiffractive collisions, which 
have larger particle production at midrapidity.  Thus, in the \pp case when 
one measures the number of midrapidity particles per event, one is 
including 75\% of the particles in the numerator and 52\% of the overall 
events in the denominator.  A similar effect will be present in \dau 
peripheral collisions, resulting in a measured invariant yield that is also 
biased to a larger value.

There is a competing bias effect in \dau.  Because the multiplicity is higher 
in \pp events with a midrapidity particle, there will be a bias in \dau 
towards higher charge in the Au-going BBC and thus towards larger 
centrality.  For peripheral events this will lead to an under counting of 
midrapidity particles, because they will migrate to more midcentral 
categorization.  This will result in a bias in peripheral \dau events to a 
smaller measured invariant yield (i.e. one that needs to be corrected up).  
In central \dau events, one will have the opposite effect (i.e. migration 
of additional midrapidity particles into this category) and the yield in 
such events will need to be corrected down.

To calculate these bias-factor corrections, we assume that a 
binary collision that produces a midrapidity particle in a given \dau event 
has a larger NBD contribution to the BBC Au-going direction charge.  From 
\pp MB and clock-trigger data, we determine this additional 
charge to be consistent with scaling both the NBD parameters $\mu$ and 
$\kappa$ up by a multiplicative factor of $1.55 \pm 0.23$.  One can also 
think of this in terms of an event with a hard scattering having a larger 
overall multiplicity in the underlying event by this factor. For this 
simple estimate, we assume the following:
\begin{enumerate} 
\item in an event with $N$ binary collisions, the one with a hard 
scattering is biased to higher multiplicity and higher trigger efficiency,
\item the increase in the BBC charge measured in \pp is applicable for this 
one binary in a \dau event, and
\item the other $N-1$ binary collisions are unaffected. 
\end{enumerate}
We then calculate the invariant yield with and without this bias in the 
Glauber+NBD framework and determine the bias-factor corrections.

\begin{figure*}[thb]
\includegraphics[width=0.99\linewidth]{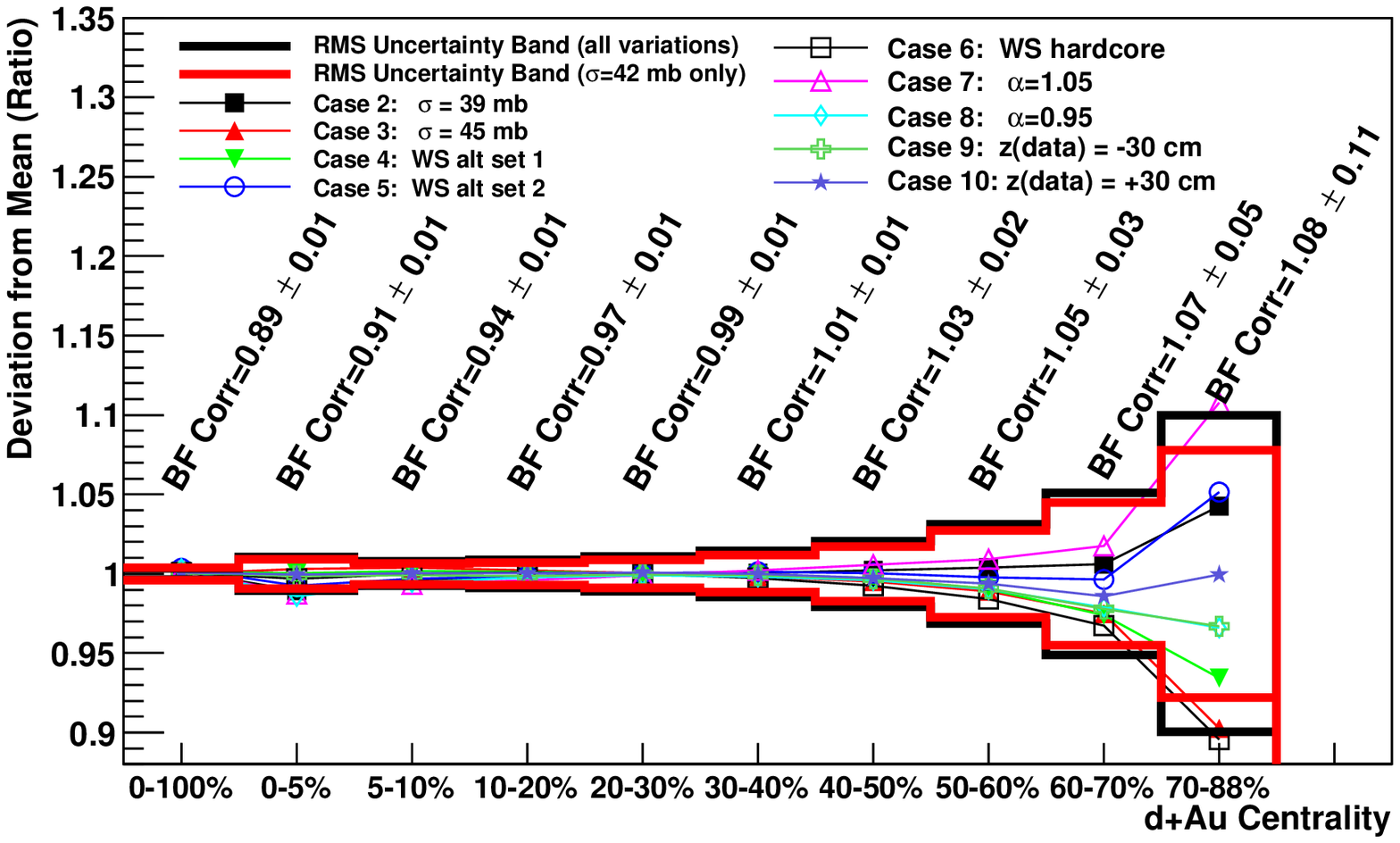}  
\caption{\label{fig:biasfactors} Multiplicative bias-factor corrections and 
their systematic uncertainties as a function of collision centrality.  
}
\end{figure*}

These bias-factor corrections can then be applied to physical quantities, 
such as \rdau using
\begin{equation}
R_{dAu} = \frac{c\ dN^{dAu}/dy}{\langle N_{\rm coll} \rangle\ dN^{pp}/dy},
\end{equation}
where $c$ is the bias-factor correction. The 0\%--100\% centrality 
integrated correction is a special case.  Because the PHENIX MB trigger 
covers only 88\% of the \dau inelastic cross section, the correction is 
used as:
\begin{equation}
R_{dAu}(0\%\text{--}100\%) = \frac{c\ dN^{dAu}/dy(0\%\text{--}88\%)}
{\langle N_{\rm coll} \rangle\ dN^{pp}/dy}.
\end{equation}
We again employ the 81 variations in the Glauber+NBD parameters and 
determine the best bias-factor corrections and their systematic 
uncertainties, as shown in Fig.~\ref{fig:biasfactors}.

For the more commonly used PHENIX four centrality bins, the 
correction factors are given in Table~\ref{tab:full4}. For the 0\%--100\% 
category, the correction factor is $0.89 \pm 0.01$.  In the most central 
\dau category, there is no trigger efficiency bias effect (i.e. such events 
always fire the trigger) and the multiplicity bias leads to an 
over-counting of particles in this category (hence a downward correction 
factor).  In the most peripheral \dau category, there is a balancing of the 
two effects.  Calculated separately, the trigger bias correction is 0.89 
and the centrality bias correction is 1.16, yielding an overall correction 
of 1.03 and a larger systematic uncertainty than the other centralities.

\subsection{Transverse-momentum dependence of bias-factor corrections}

The above calculation of the bias-factor corrections was based essentially 
on two input values: 
\begin{enumerate}
\item the 75\% probability of the BBC-MB trigger to fire when 
a particle is detected at midrapidity in \pp collisions, and 
\item the increase in BBC multiplicity by 1.55 when a particle is detected 
at midrapidity in conjunction with a BBC-MB trigger firing.  
\end{enumerate}
Earlier PHENIX studies indicated that these two \pp values 
were independent of the \pt of the midrapidity particle and 
of particle types (including $\pi^{0}$, unidentified $h^{\pm}$, $J/\psi$) 
within uncertainties.  However, these studies were limited to transverse 
momentum values less than 9~GeV/$c$~\cite{Adler:2003qs}.

\begin{figure*}[thb]
\includegraphics[width=0.99\linewidth]{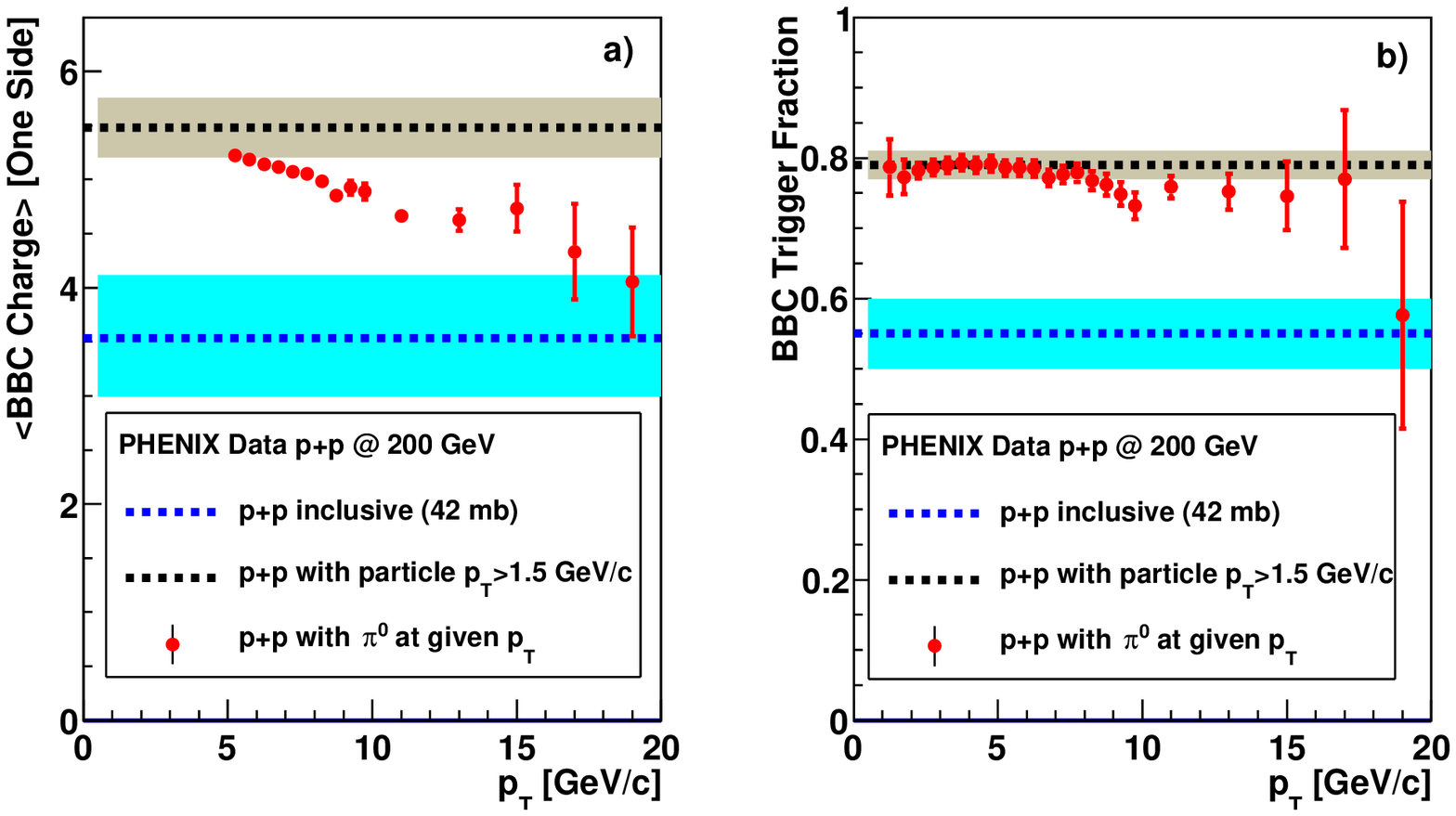}  
\caption{\label{fig:ppdatabias} 
(a) Shown as a dashed blue line is the mean BBC charge on one side for 
inclusive \pp interactions, corresponding to $\sigma_{inelastic} = 42$ mb.  
The systematic uncertainty is shown as a band.  Shown as a dashed black 
line is the mean BBC charge on one side when a midrapidity charged hadron 
with \pt$>1.5$~GeV/$c$ is in coincidence.  The red points are the mean 
BBC charge when there is a $\pi^{0}$ measured with a \pt as indicated 
by the x-axis.  (b) The same quantities as before, but for the 
fraction of interactions that fire the BBC coincidence trigger.  The 
bias-factor corrections were originally calculated in 2003 with inputs that 
the BBC trigger fires on $52 \pm 4$\% of inclusive inelastic \pp interactions 
corresponding to 42 mb and $75\pm3$\% of inelastic \pp interactions with 
a charged hadron of $p_T>1.5$~GeV/$c$.  For the 2008 analysis, the BBC 
thresholds were adjusted and the most accurate determination indicates that 
the BBC trigger fires on $55 \pm5 $\% of inclusive inelastic \pp interactions 
corresponding to 42 mb and $79 \pm 2$\% of inelastic \pp interactions with 
a charged hadron of $p_{T} > 1.5$~GeV/$c$, as shown above.  The $\pi^{0}$ 
points shown above are from data taken with these thresholds adjusted.  
These differences result in negligible changes to the bias-factor 
correction values used.  
}
\end{figure*}

High statistics results for neutral pions indicate a change in these values 
at much higher \pt, as shown in Fig.~\ref{fig:ppdatabias}.  These values 
are determined from \pp data taken in 2006 utilizing a photon trigger with 
and without the coincidence on the BBC-MB trigger. There is 
always an increase in the mean multiplicity in events with a neutral pion 
compared to MB \pp events; however, that value decreases by 
approximately 20\% for particles near \pt$\approx 15-20$~GeV/$c$ 
compared with lower \pt, as shown in the left panel of 
Fig.~\ref{fig:ppdatabias}.  The trigger efficiency shows a very slight 
decrease for the highest \pt measured, as shown in the right panel of 
Fig.~\ref{fig:ppdatabias}.

As a simple estimate of the change in bias-factor corrections for these 
high $p_T$ particle invariant yields, we can repeat the procedure from the 
previous section replacing the values of 75\% with 70\% and the increased 
multiplicity factor of 1.55 with 1.25, as overestimates for neutral pions 
with $p_{T} \approx 15$~GeV/$c$.  Changing these factors results in a bias 
factor correction for central (i.e. 0\%--20\%) \dau events of 0.97 
(compared with the previous value of $0.94 \pm 0.01$).  This result makes 
sense in that there is slightly less centrality shifting bias, because the 
increased multiplicity is reduced.  Also, because there are a large number of 
binary collisions (of order 15), the bias from just one binary collision 
leaving the other $N-1$ unmodified, yields only a small change.  In 
considering the peripheral category (i.e. 60\%--88\% centrality bin), it is 
interesting to look first at the two bias contributions separately.  The 
trigger part of the bias-factor correction is now 1.07 (compared with the 
previous value of 1.16) and the centrality shifting bias correction is now 
0.90 (compared with the previous value of 0.89).  The overall combined bias 
correction becomes 0.96 (compared with the previous value of $1.03 \pm 
0.06$).  These results are slightly outside of the RMS systematic 
uncertainties quoted on these bias-factor corrections.

One might hypothesize about the origin of this $p_T$ dependence and posit 
that it relates to using up more energy at midrapidity thus yielding a 
decrease in particles at backward rapidity, or a change in the rapidity 
distribution itself. It is also unclear that the other $N-1$ binary 
collisions are uncorrelated with the process in the one binary collision 
producing the midrapidity high $p_T$ particle.  This motivates a full 
{\sc hijing} MC study where we know the true invariant yields and can determine 
(albeit in a model-dependent way) the actual bias-factor corrections due to 
auto-correlations.  The goal of this {\sc hijing} study is not to correct the 
experimental data in a model-dependent way, but rather to gain some insight 
into the centrality and bias correction method.

\section{{\sc hijing} Study}
\label{sec:hijing}

The {\sc hijing} MC generator~\cite{Gyulassy:1994ew} has been 
established as a useful tool for the study of hard scattering processes and 
the underlying event in \pp and $A$+$A$ collisions over a wide range of 
collision energies.

\subsection{Centrality Bias-Factor Corrections for \dau at \sqsn=~200~GeV}  

Prior to carrying out the bias-factor study, we need to define a set of 
selection cuts to make a close comparison with the experimental results.  
First, we model the PHENIX BBC response and trigger selection by examining 
the number of particles within the same pseudorapidity acceptance \bbceta.  
A full {\sc geant} simulation of the BBC response on {\sc hijing} events is 
computationally prohibitive, because we examine tens of billions of {\sc hijing} 
events to study the \pt dependence of the bias-factor corrections.

The PHENIX MB trigger is modeled requiring at least one particle in each of 
the BBC regions. We find that for \pp collisions at \sqsn=~200~GeV, the 
percentage of {\sc hijing} events satisfying the PHENIX MB trigger 
requirement is 48\% compared with the experimentally measured value of 52 
$\pm$ 4\%.  For {\sc hijing} simulated \dau collisions, the trigger 
requirement is met by 83\% of events compared with the previously quoted 
value of 88 $\pm$ 4\%. For \dau collisions, we divide the simulated {\sc hijing} 
BBC multiplicity distribution into centrality selections following the same 
procedure used on experimental data.  In the {\sc hijing} case, we can examine 
the centrality selected events for the true number of binary collisions.  
The mean and root-mean-square (RMS) \mNcoll values from {\sc hijing} are given in 
Table~\ref{tab:hijing_ncoll}.  These values are in reasonable agreement 
with those given earlier as determined from experimental data and the 
Glauber+NBD fit. The slight difference in the 60\%--88\% category is 
potentially due to the differences in trigger efficiencies previously 
noted. The RMS values are somewhat smaller from {\sc hijing}, which may be due to 
the lack of a complete simulation of the BBC detector response.

\begin{table}
\caption{Mean \Ncoll and RMS values for each centrality}
\begin{ruledtabular} \begin{tabular}{ c@{\hskip 0.2in}  c  c@{\hskip 0.2in}  c c }
 & \multicolumn{2}{c}{\textbf{{\sc hijing}}} & \multicolumn{2}{c}{\textbf{Glauber+NBD}}\\
Centrality & \mNcoll & RMS &  \mNcoll & RMS\\
\hline 
 0\%--20\% & 15.0 & 4.1 & 15.1 $\pm$ 1.0 & 4.9\\
 20\%--40\% & 10.1 & 3.5 & 10.2 $\pm$ 0.7 & 4.2\\
 40\%--60\% & 6.3 & 3.0 & 6.6 $\pm$ 0.4 & 3.6\\
 60\%--88\% & 2.8 & 2.0 & 3.2 $\pm$ 0.2 & 2.3\\
  \end{tabular} \end{ruledtabular}
\label{tab:hijing_ncoll}
\end{table}

In {\sc hijing} the requirement of a midrapidity (i.e. $|\eta|<0.35$) particle 
with \pt $>$ 1~GeV/$c$ in \pp collisions increases the BBC multiplicity by 
a factor of 1.62 (compared with 1.55 as measured with experimental data) 
and increases the probability to satisfy the BBC-MB trigger requirement to 
62\% (compared with 75 $\pm$ 3\% as measured with experimental data).  The 
trigger difference may be due to the specific handling of single and double 
diffractive events within {\sc hijing}.  With these differences kept in mind, we 
proceed to calculate the bias-factor corrections within {\sc hijing}.

First, we separate {\sc hijing} \dau events into four centrality selections 
(0\%--20\%, 20\%--40\%, 40\%--60\%, 60\%--88\%) by the simulated BBC 
multiplicity in the Au-going direction. Using the generator-level truth 
information for these events, we determine the distribution of \Ncoll for 
each centrality. We then calculate the yield of particles at midrapidity 
per event within each centrality selection, as is done with the 
experimental data.  We refer to this as the {\sc hijing} ``measured'' yield. 
Separately, using the truth information on the number of binary collisions, 
we sort all {\sc hijing} events into the four centrality bins to exactly match 
the \Ncoll distributions determined from the ``measured'' selection.  We 
calculate the yield of particles at midrapidity per event using this truth 
information, not the multiplicity of the event.  We refer to this yield as 
the {\sc hijing} ``truth'' yield. The only difference between the {\sc hijing} 
``measured'' and ``truth'' yields results from the auto-correlation between 
the midrapidity particle production and the multiplicity measured in the 
Au-going direction.  Therefore, the ratio of ``truth'' to ``measured'' is 
exactly the bias correction factor.  The {\sc hijing} bias-factor corrections for 
midrapidity particles with \pt $>$ 1~GeV/$c$ are shown in 
Table~\ref{tab:hijing_bias_factors}.  The uncertainties shown are 
statistical only.  For comparison, we again include the experimentally 
determined bias-factor corrections from the Glauber+NBD procedure.  The 
correction factors are in agreement within uncertainties with those derived 
from Glauber+NBD


\begin{table*}
\caption{\label{tab:hijing_bias_factors}
Mean bias-factor corrections as a function of \pt for each 
centrality as calculated with {\sc hijing}, and comparison with reference 
Glauber+NBD values.
}
\begin{ruledtabular} \begin{tabular}{cccccc}
           &             & {\sc hijing}   & {\sc hijing}      & {\sc hijing}   & {\sc hijing}\\
Centrality & Glauber+NBD & $1\le\pt<5$    & $5\le\pt<10$      & $10\le\pt<15$  & $15\le\pt<20$\\
\hline
0\%--20\%  & $0.94\pm 0.01$ & $0.951\pm 0.001$& $0.962\pm 0.001$ & $1.000\pm 0.005$ & $1.038 \pm 0.020$\\
20\%--40\% & $1.00\pm 0.01$ & $0.996\pm 0.001$& $1.008\pm 0.001$ & $1.010\pm 0.006$ & $0.996 \pm 0.021$\\
40\%--60\% & $1.03\pm 0.02$ & $1.010\pm 0.001$& $1.022\pm 0.001$ & $1.019\pm 0.007$ & $1.005 \pm 0.025$\\
60\%--88\% & $1.03\pm 0.06$ & $1.030\pm 0.001$& $1.026\pm 0.001$ & $0.999\pm 0.008$ & $0.991 \pm 0.030$\\
\end{tabular} \end{ruledtabular}
\end{table*}

\begin{figure}[thb]
\includegraphics[width=1.0\linewidth]{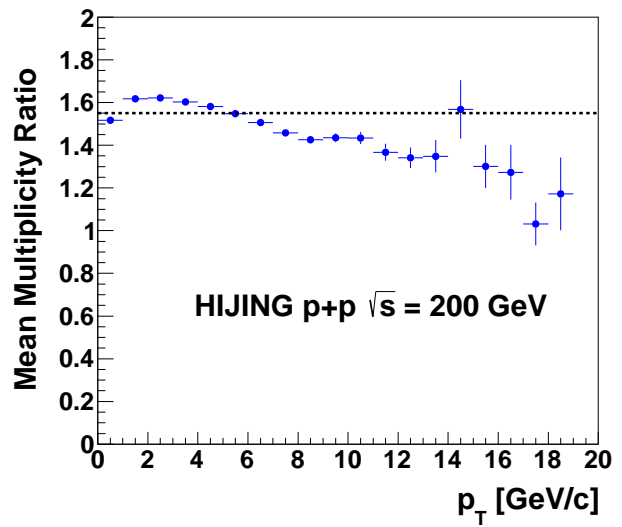}  
\caption{\label{fig:mBBCS_ratio_RHIC}
The ratio of the mean multiplicity at $-3.9<\eta<-3.0$ in triggered events 
with a particle with a given \pt produced at midrapidity to all inelastic 
\pp collisions from {\sc hijing} at \sqsn=~200~GeV. The 
dashed line at 
1.55 represents the mean reference value measured in data.
}
\end{figure}

\begin{figure}[thb]
\includegraphics[width=1.0\linewidth]{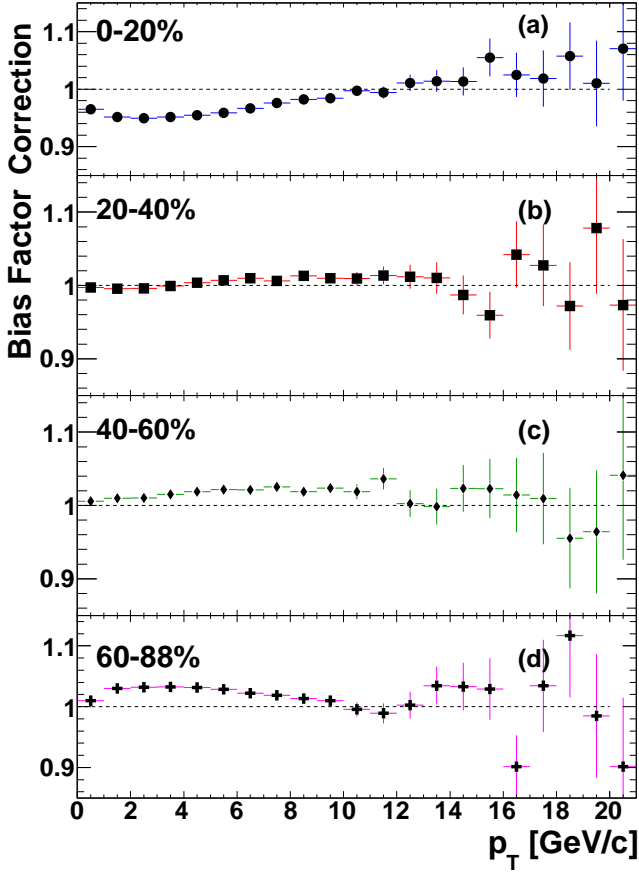}  
\caption{\label{fig:hijing_bias_factors_RHIC} 
Bias-factor corrections as a function of \pt for {\sc hijing} $d$$+$Au events at 
\sqsn=~200~GeV.
}
\end{figure}

With a large statistical sample of {\sc hijing} \pp and \dau events, we 
can also examine the \pt dependence of these bias-factor corrections.  
Figure~\ref{fig:mBBCS_ratio_RHIC} shows the {\sc hijing} \pp mean 
multiplicity in the BBC for events which contain a particle at 
midrapidity with a given \pt, as a ratio relative to all inelastic \pp 
collisions.  The results indicate a decrease of the multiplicity for 
higher \pt particles at midrapidity, in qualitative agreement with that 
shown earlier in \pp experimental data.  Following the same procedure 
outlined above for determining the {\sc hijing} ``measured'' and 
``truth'' yields, as well as the same centrality definitions, we 
calculate the bias-factor corrections as a function of the \pt of the 
midrapidity particle. The results are shown in 
Fig.~\ref{fig:hijing_bias_factors_RHIC}.  These values are integrated 
over wider \pt selections and are tabulated in 
Table~\ref{tab:hijing_bias_factors}. These bias correction factors vary 
by less than 5\% (10\%) up to $p_{T}$ = 10 (20)~GeV/$c$ in all four 
centrality categories. There is a slight increase in the bias-factor 
correction in central events in agreement and a slight decrease in the 
bias-factor correction in peripheral events, both in agreement with our 
simple estimate from the experimental \pp data values.  Though these 
results are {\sc hijing} model specific, the agreement of the bias 
correction factors and their very modest \pt dependence are noteworthy.  
There are other models that make different assumptions regarding the 
relationship between binary collisions and geometry and therefore may 
give different results, for example see Ref.~\cite{Alvioli:2013vk}.

\subsection{Centrality Bias-Factor Corrections for \ppb at \sqsn=~5.02~TeV}  

\begin{figure}[thb]
\includegraphics[width=1.0\linewidth]{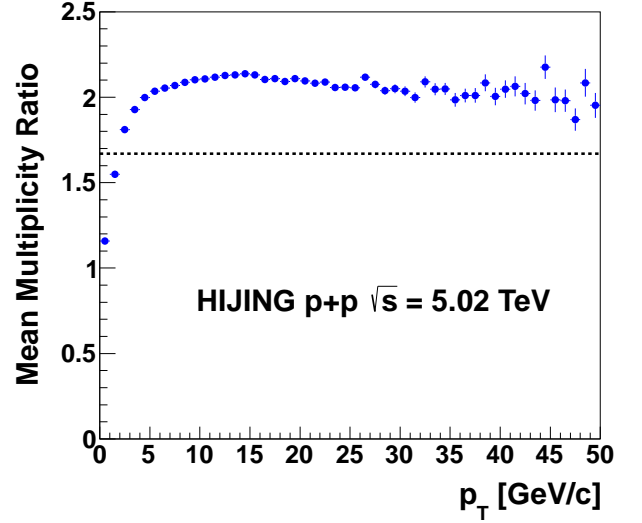} 
\caption{\label{fig:mBBCS_ratio_LHC} 
The ratio of the mean multiplicity at $-4.9<\eta<-3.1$ in triggered events 
with a particle with a given \pt produced at midrapidity to all inelastic 
\pp collisions from {\sc hijing} at \sqsn=~5.02~TeV. The dashed line 
corresponds to the inclusive mean multiplicity ratio.
}
\end{figure}

Given the recent \ppb collision data at \sqsn=~5.02~TeV at the LHC, 
it is interesting to apply the identical procedure using 
{\sc hijing} at this higher energy.  In this case, we use the particle 
multiplicity within $-4.9 < \eta < -3.1$ as the simulated detector 
acceptance in the Pb-going direction for determining centrality quantiles.  
From {\sc hijing} \pp events at 5.02~TeV, we find that the presence of a 
midrapidity particle increases the Pb-going multiplicity by a factor of 
approximately 1.67 and with a significant dependence on the \pt of the 
midrapidity particle, as shown in Fig.~\ref{fig:mBBCS_ratio_LHC}.

\begin{figure}[!thb]
\includegraphics[width=1.0\linewidth]{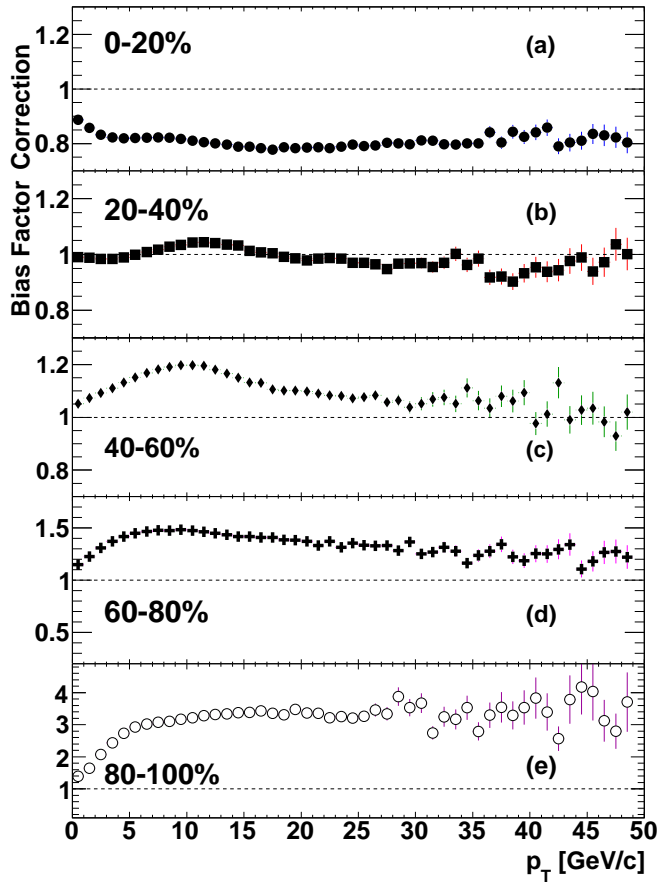}  
\caption{\label{fig:bias_factors_LHC} 
Bias-factor corrections as a function of \pt for {\sc hijing} \ppb events at 
\sqsn=~5.02~TeV. Correction factors for each centrality are plotted with 
different vertical scales to better illustrate the magnitude of the effect.
}
\end{figure}

The requirement of one particle in the forward and backward acceptance for 
these {\sc hijing} \ppb events has a nearly 100\% efficiency.  Thus, we divide 
the \ppb events into five centrality categories 0\%--20\%, 20\%--40\%, 
40\%--60\%, 60\%--80\%, and 80\%--100\%.  The bias-factor corrections thus 
are expected to only include the centrality migration effect and no effect 
from the trigger auto-correlation bias.  The resulting bias-factor 
corrections as a function of \pt are shown in 
Fig.~\ref{fig:bias_factors_LHC}.  The {\sc hijing} calculations indicate 
very large correction factors in the most peripheral selection and with a 
substantial \pt dependence, particularly over the range $1<\pt<10$~GeV/$c$.  
This means that the {\sc hijing} ``measured'' yield at \pt = 5~GeV/$c$ 
would be more than a factor of two lower than the truth value.

\subsection{{\sc hijing} Discussion}

\begin{figure*}[!thb]
\includegraphics[width=0.45\linewidth]{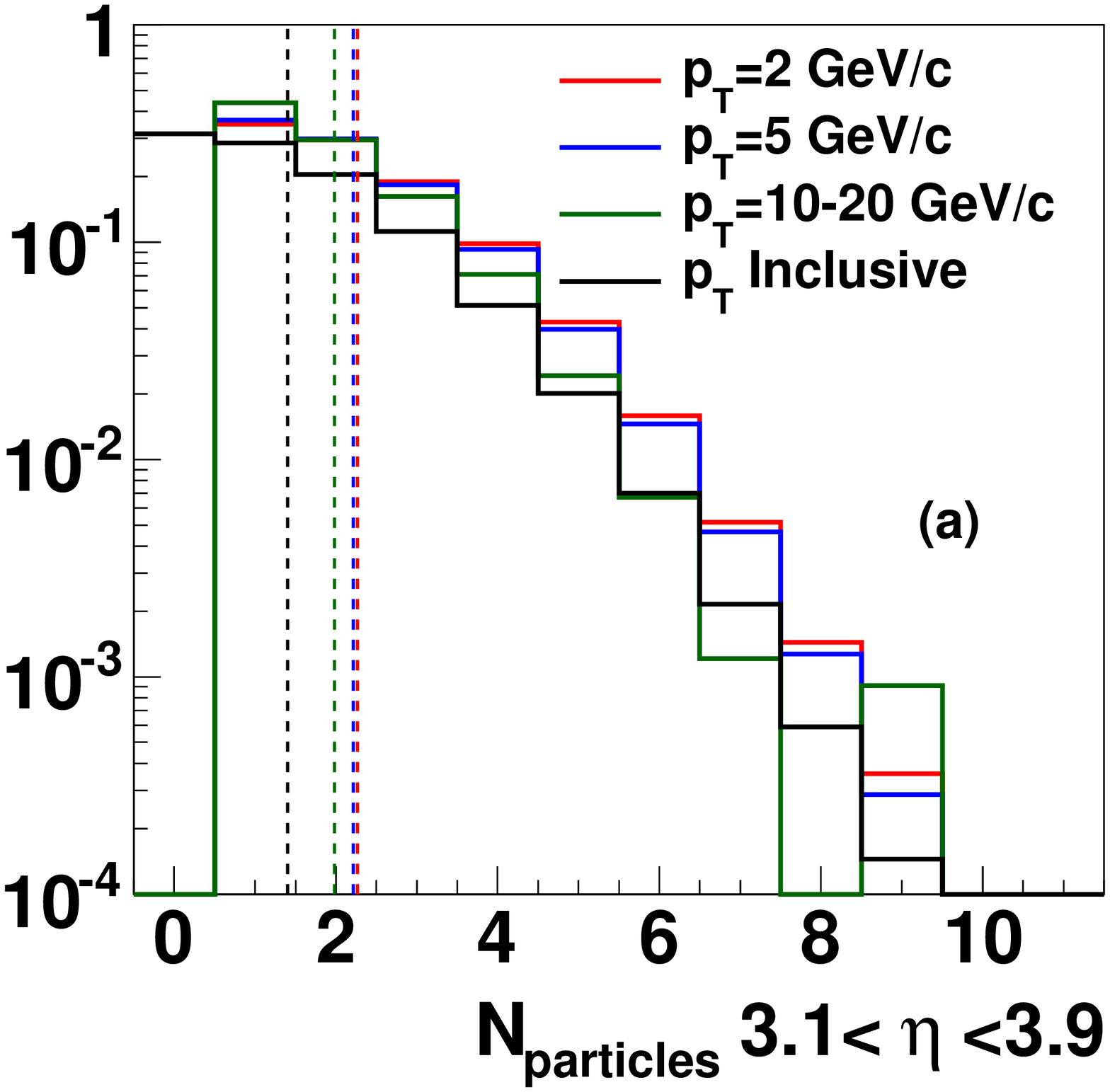}  
\includegraphics[width=0.45\linewidth]{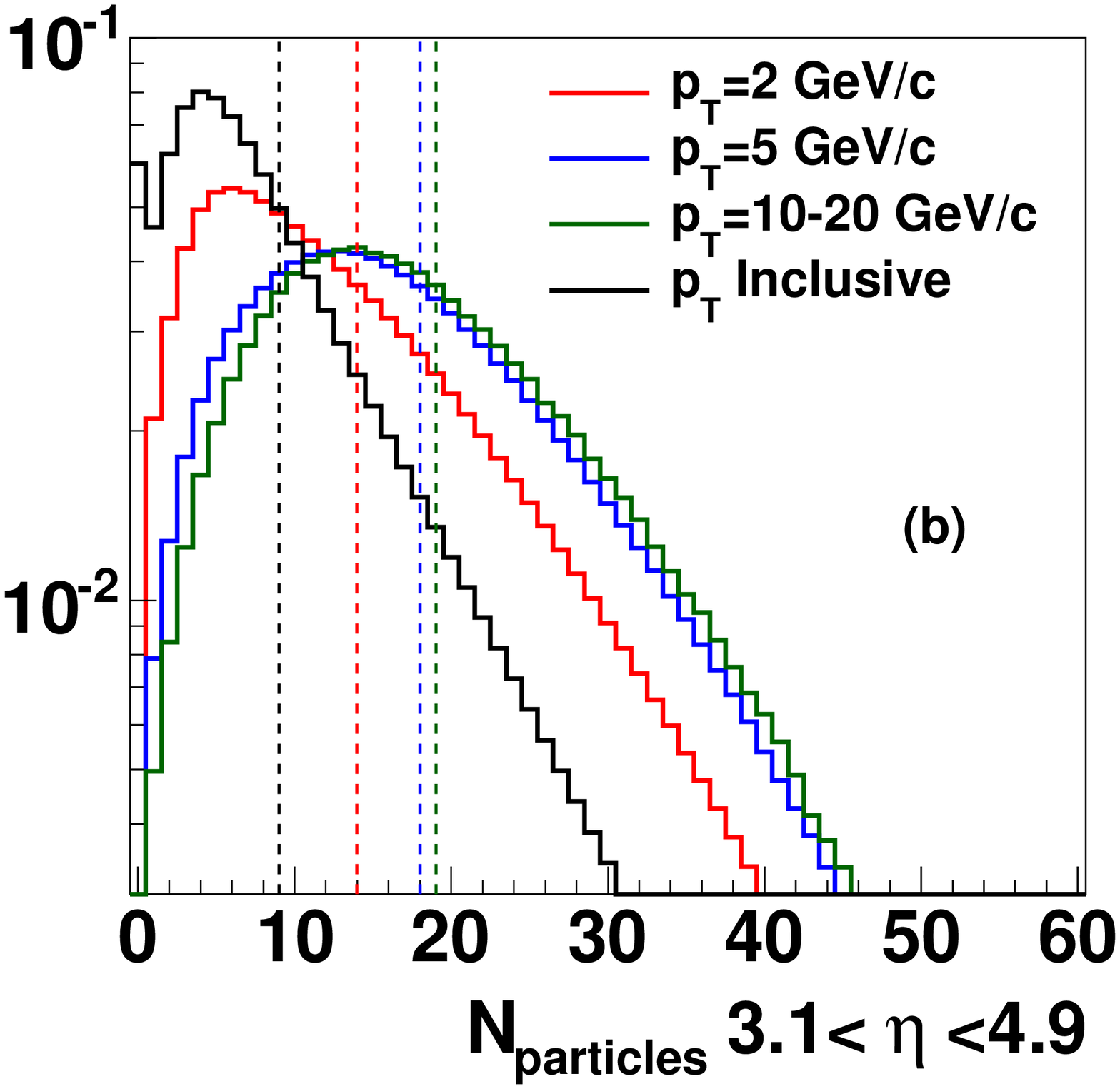}  
\caption{\label{fig:pp_multiplicity_dist_LHCRHIC} 
Multiplicity distribution at $-4.9<\eta<-3.0$ for {\sc hijing} \pp events at 
(left) \sqs=~200~GeV at RHIC and (right) \sqs=~5.02~TeV at the LHC. 
Dashed lines indicate the mean values of each distribution.  
}
\end{figure*}

The bias factors extracted from {\sc hijing} in \ppb collisions at 
\sqsn=~5.02~TeV are an order of magnitude larger than those in \dau 
collisions at \sqsn=~200~GeV.  When comparing the \ppb and \dau results, 
it should be noted that in the most peripheral class, the \dau case only 
extends down to 88\% due to the trigger efficiency and part of the 
centrality migration bias is canceled by the trigger bias. 
Figure~\ref{fig:pp_multiplicity_dist_LHCRHIC} compares the {\sc hijing} 
\pp multiplicity distribution in the backward acceptance for different 
selections on the \pt of a midrapidity particle.  One observes only a 
modest dependence on the \pt of the midrapidity particle for RHIC 
energies, and a large dependence for LHC energies.  This auto-correlation 
directly translates into the large calculated bias-factor corrections.

The {\sc hijing} results follow the same trends previously observed in 
$p$$+$$\overline{p}$ collisions at the Tevatron from \sqs=~300~GeV to 
1.96~TeV~\cite{Abe:1993rv,Abe:1997xk,Abazov:2009gc} and in \pp 
collisions at the LHC from \sqs~=900~GeV to 
7.0~TeV~\cite{Aaij:2011yc,Aaij:2012dz,Aad:2013bjm}, as detailed 
in~\cite{Seymour:2013qka}. At the highest LHC energies, the increased 
multiplicity of the underlying event as the particle $p_T$ value 
increases from 1 to 10~GeV/$c$, is well described in terms of multiparton 
interactions (MPI)~\cite{Seymour:2013qka}. At the lowest Tevatron energy 
of 300~GeV, the underlying event is relatively unchanged for $p_T$ values 
from 2~GeV/$c$ and above due to the much lower hard scattering cross 
section and thus the smaller influence from multiparton interactions.

The effect of MPI at RHIC and LHC energies can be investigated within 
{\sc hijing}. According to {\sc hijing}, the mean number of hard scatterings per 
nucleon-nucleon binary collision is 0.24 in \dau at 200~GeV, but 1.36 in 
\ppb at 5.02~TeV. This highlights the strong $\sqrt{s}$ dependence of the 
effect.

\begin{figure}[thb]
\includegraphics[width=1.0\linewidth]{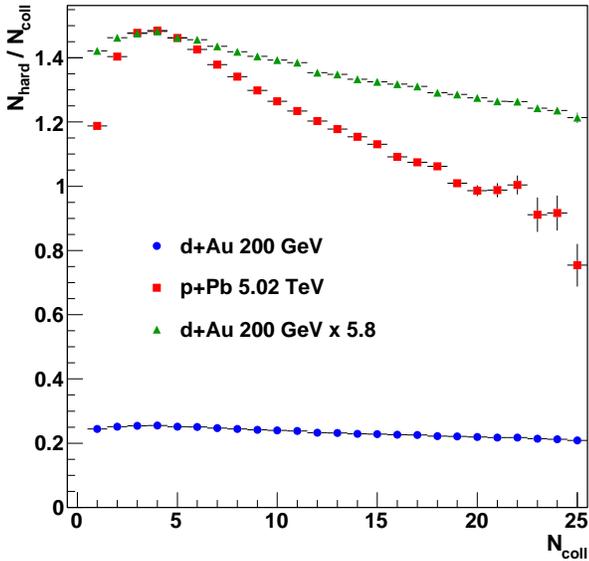}
\caption{\label{fig:MPI_dAupPb} 
A comparison of the MPI effect in {\sc hijing} for \dau and \ppb collisions at 
200~GeV and 5.02~TeV respectively. Plotted is the number of hard 
scatterings ($N_{\rm hard}$) divided by the number of nucleon-nucleon 
collisions (\Ncoll) as a function of \Ncoll.
}
\end{figure}

The {\sc hijing} calculations also include another effect that will 
result in deviations from binary scaling, but is not an auto-correlation 
effect. Figure~\ref{fig:MPI_dAupPb} shows the number of hard scatterings 
per nucleon-nucleon collision as a function of \Ncoll. Peripheral \pda 
events have individual binary collisions geometrically biased towards 
larger impact parameters, i.e. with less overlap between the nucleons, 
within the constraint $b<\sqrt{\sigma_{NN}/\pi}$. {\sc hijing} has a 
geometric overlap dependence to the hard scattering probability. This 
effect is significantly larger in \ppb collisions at 
\sqsn=~5.02~TeV~\cite{fortheALICE:2013xra}. Additionally the decrease for 
more central events may be an energy conservation effect, which is also 
smaller for \dau collisions at \sqsn=~200~GeV, because the binary scatters 
are split between two projectile nucleons. It is unclear that one should 
attempt to correct for this in constructing \rpda. In either case the 
effects are small for \dau at \sqsn=~200~GeV.

The results of the {\sc hijing} study shows that the bias effects are small with 
little \pt dependence in \dau at \sqsn=~200~GeV, and that these bias 
factors are primarily due to the trigger bias toward nondiffractive collisions. 
However, in \ppb collisions at \sqsn=~5.02~TeV, the bias factors are 
large 
and dominated by the effects of multiparton interactions.

\section{Summary}
\label{sec:summary}

We have presented a detailed description of centrality determination in 
\pda collisions. This method has been utilized by PHENIX in the analysis of 
\dau collision data recorded in 2008. Using a Glauber-MC calculation coupled 
with a simulation of the charge deposited in the Au-going BBC we are able 
to determine geometric quantities associated with different centrality 
selections. Using this model, we also calculate the fraction of events in 
each centrality class with a spectator neutron, and find good agreement 
with the measured data. Utilizing the same formalism we also describe a 
calculation of the bias-factor corrections associated with this centrality 
determination. Using {\sc hijing}, we present a study of the \pt dependence of 
these bias-factor corrections. We find that they exhibit a modest \pt 
dependence at \sqsn=~200~GeV, in agreement with those derived with the 
Glauber-model and used in earlier PHENIX publications. Repeating the {\sc hijing} 
study for \ppb collisions at \sqsn=~5.02~TeV, we find significantly 
larger centrality bias factors, exhibiting a strong \pt dependence that may 
result in part from the much higher contribution from multiparton 
interactions. Additional experimental checks are needed, at both energies, 
of $p$$+$A collisions with different $A$ to further study these effects.


\section*{ACKNOWLEDGMENTS}   

We thank the staff of the Collider-Accelerator and Physics
Departments at Brookhaven National Laboratory and the staff of
the other PHENIX participating institutions for their vital
contributions.  We acknowledge support from the 
Office of Nuclear Physics in the
Office of Science of the Department of Energy, the
National Science Foundation, 
Abilene Christian University Research Council, 
Research Foundation of SUNY, and 
Dean of the College of Arts and Sciences, Vanderbilt University (U.S.A),
Ministry of Education, Culture, Sports, Science, and Technology
and the Japan Society for the Promotion of Science (Japan),
Conselho Nacional de Desenvolvimento Cient\'{\i}fico e
Tecnol{\'o}gico and Funda\c c{\~a}o de Amparo {\`a} Pesquisa do
Estado de S{\~a}o Paulo (Brazil),
Natural Science Foundation of China (P.~R.~China),
Ministry of Education, Youth and Sports (Czech Republic),
Centre National de la Recherche Scientifique, Commissariat
{\`a} l'{\'E}nergie Atomique, and Institut National de Physique
Nucl{\'e}aire et de Physique des Particules (France),
Bundesministerium f\"ur Bildung und Forschung, Deutscher
Akademischer Austausch Dienst, and Alexander von Humboldt Stiftung (Germany),
Hungarian National Science Fund, OTKA (Hungary), 
Department of Atomic Energy and Department of Science and Technology (India), 
Israel Science Foundation (Israel), 
National Research Foundation and WCU program of the 
Ministry Education Science and Technology (Korea),
Physics Department, Lahore University of Management Sciences (Pakistan),
Ministry of Education and Science, Russian Academy of Sciences,
Federal Agency of Atomic Energy (Russia),
VR and Wallenberg Foundation (Sweden), 
the U.S. Civilian Research and Development Foundation for the
Independent States of the Former Soviet Union, 
the US-Hungarian Fulbright Foundation for Educational Exchange,
and the US-Israel Binational Science Foundation.



\end{document}